\documentclass[reprint,groupedaddress,superscriptaddress,aps,pre]{revtex4-2}

\usepackage[dvipsnames]{xcolor}
\definecolor{linkcolor}{rgb}{0.3,0.3,1.0} %hyperlink
\usepackage[pdftex,colorlinks=true, linkcolor= linkcolor, citecolor= linkcolor, urlcolor= linkcolor, hyperindex=true,hyperfigures=true]{hyperref} %hyperlink
\usepackage[utf8]{inputenc}

\usepackage{amssymb,amsmath,amsfonts}
\usepackage{graphicx}
\usepackage{float}
\usepackage{xspace}
\usepackage{bm}
\usepackage{comment}
\usepackage{balance}

\newcommand{\eg}{\emph{e.g.}\xspace}

\begin{document}
%%%
%\title{Random Sequential Adsorption of DLA Clusters on Square Lattice: Effect of Polydispersity on Jamming and Percolation}
%\title{Jamming states of fractal DLA clusters in random sequential adsorption}
\title{Jamming states in random sequential adsorption of diffusion-limited aggregates}
%%%

%%%
\author{Fahad Puthalath}
\affiliation{Institut f\"ur Frontier Materials auf der Erde und im Weltraum , Deutsches Zentrum f\"ur Luft- und Raumfahrt (DLR), 51147 K\"oln, Germany}
\affiliation{Institut f\"ur Theoretische Physik, Universität zu K\"oln, Z\"ulpicher Strasse 77, 50937 K\"oln, Germany}
\author{Dipanjan Mandal}
\affiliation{Department of Physical and Chemical Sciences, University of L’Aquila, Via Vetoio, 67100 L’Aquila, Italy}
\author{Sumanta Kundu}
\email{skundu@sissa.it}
 \affiliation{Dipartimento di Fisica, Università di Napoli Federico II, and INFN Napoli, Complesso Universitario di Monte
Sant’Angelo, 80126 Napoli (IT)}
\affiliation{Scuola Internazionale Superiore di Studi Avanzati (SISSA), via Bonomea 265,  34136 Trieste (IT)}

\date{\today}

\begin{abstract}
Motivated by the ubiquity of ramified fractal deposits in nature and engineered systems, we investigate the irreversible adsorption of diffusion-limited aggregation (DLA) clusters on a square lattice. We study the role of cluster shape diversity on jamming properties of the system by systematically controlling the number of distinct shapes used across and within realizations, encompassing both monodisperse and polydisperse model variants. Our large-scale simulations over a broad range of cluster sizes $2\leqslant k\leqslant4096$ show that the jamming density decreases with cluster size as a power-law $p_j(k)-p_j^\infty\sim k^{-\alpha}$. Both $\alpha$ and $p_j^\infty$ are found to depend on the degree of shape diversity, with $\alpha$ ranging from $0.374(2)$ to $0.417(1)$. It is observed that increasing shape polydispersity promotes denser packing. Importantly, the fluctuations of the jamming density exhibit distinct scaling behavior: $\sigma(L)\sim1/L$ for a fixed pool of cluster shape(s), but remain $L$-independent when the pool of shape(s) is refreshed across different realizations. Furthermore, our results demonstrate that the differences between the model variants systematically diminish with increasing $k$ and are expected to vanish as $k\to\infty$ due to the statistical self-similarity of the DLA clusters.
\end{abstract}

\maketitle

\section{Introduction}
Adsorption or deposition of objects onto substrates is ubiquitous in nature, ranging from the adsorption of protein molecules on biological membranes to the fabrication of nanostructured thin films for technological applications
~\cite{
%Biollogy-in-order-of-years
Finegold1979,Balaz1984,
%(nano-powders)-in-order-of-years
Schwager2008,
%nano-thinfilms-in-order-of-years
Pablo1999,Plawsky2009,Joshi2016}. A broad class of these deposition processes leads to the formation of non-overlapping monolayer, close-packed jamming states, whose structural properties depend on the geometry of the depositing objects. Although the dependence of jamming properties on object size and shape is well understood for regular geometries, it remains far less explored for  objects having irregular shapes. This opens up the fundamental question of how irregular$-$\eg, fractal-like objects affect the properties of the jamming states.

A widely used theoretical framework for studying such irreversible deposition and jamming is the random sequential adsorption (RSA) model~\cite{Evans1983,Evans1993,Schaaf2000,Talbrot2000,Kubala2022,Feng2025}. In RSA model, objects are deposited sequentially at randomly selected positions on a substrate and absorbed if they do not overlap with any previously adsorbed objects; otherwise, the deposition attempt is rejected, and again a new empty position is selected at random for next deposition attempt. The process terminates when no additional object can be accommodated on the available empty space, resulting in a jamming state.

Among regular shapes, RSA of line segments~\cite{Manna1991,Bonnnier1994,Kondrat2001,Lebovka2011,Koza2025}, squares~\cite{Nakamura1986,Borosilov1991,Ramirez-Square}, rectangles~\cite{Ziff1989,Lebovka2020,Petrone2021}, disks~\cite{Feder1980,Onoda1986}, ellipses~\cite{Sherwood1990,Viot1992,Ciesla2016}, and polydisperse mixtures with prescribed size distributions~\cite{Meakin1992,Brilliantov1996,Hart2016,Wagaskar2020,Kundu2022} has been extensively investigated in both continuum and lattice systems. These studies have established that the shape, anisotropy, and the orientational alignment of the objects strongly influence the morphology of jamming states.

Nevertheless, many naturally occurring deposits are highly irregular and exhibit ramified, fractal-like patterns~\cite{Mandelbrot1982,Feder1988,Ball2009,Kozicki2021,Morris2026}. Examples include mineral deposits in fractured rocks~\cite{Vicsek1991}, sedimentary rock deposits formed by erosion and their transport in desert drainage networks~\cite{Ignacio1992}, and snowflakes~\cite{Kozicki2021}. Similar fractal morphologies also emerge in nonequilibrium thin-film growth experiments utilizing cluster deposition techniques, where atoms or clusters of atoms deposited onto a substrate produce ramified aggregates resembling diffusion-limited aggregation (DLA) clusters~\cite{Pablo1999,Cahuzac2003,Lando2006,Veronika2011,Fairbanks2011}. The formation of dendrite structures during metallic melt solidification is another example~\cite{Becker2019,Becker2023}. Despite the ubiquity of such fractal deposits, it is not yet understood how their fractal morphology influences the jamming properties.

The closest existing study considers the RSA of self-avoiding walk (SAW) chains as a model of polymer adsorption~\cite{Wang1996,Budinski1996,Adamczyk2008,Petkovic2008,Ramirez2023}. However, the relatively short chain lengths investigated in that study limit the emergence of their fractal characteristics. In fact, statistically reliable results for SAW chains have been reported only for chain lengths up to $k\leqslant 15$ on different lattices~\cite{Ramirez2023}. A systematic study of the RSA of large fractal-like clusters is still lacking.

Motivated by this and by the evidence of DLA-like cluster deposits in natural and synthetic systems, we investigate the RSA of fractal-like clusters following the DLA process on a square lattice over a broad range of cluster sizes $2\leqslant k \leqslant 4096$. Notably, DLA clusters exhibit a pronounced ramified branching morphology arising from screening effects during growth~\cite{Witten1981} and are characterized by a fractal dimension $D_f=(D^2+1)/(D+1)$ in D-dimensions~\cite{Muthukumar1983,Tokuyama1984}. Experimental observations also support these values: the growth of bacterial colonies of {\it Bacillus subtilis} on two-dimensional plates exhibit DLA structures with $D_f=1.73\pm0.02$~\cite{1990_matsushita_phyA,Tronnolone2018}. Similarly, in three dimensions, electrolytic growth of copper on a point-like cathode from an aqueous solution produces fractal structures with $D_f=2.43\pm0.03$~\cite{Brady1984}.

In this paper, we ask three fundamental questions associated with jamming: (i) Do different DLA cluster shapes of a sufficiently large size $k$ produce substantially different jamming densities, or does the underlying statistical self-similarity suppress the resulting variability in jamming densities? (ii) Does the additional randomness from shape-to-shape variations affect the scaling of the fluctuations of jamming-density with system size when different cluster shapes are used across runs, as opposed to a fixed cluster shape? (iii) Does the cluster shape polydispersity lead to more efficient packing?

%most existing studies have focused on compact or weakly anisotropic objects. In contrast, many physical systems involve the deposition of highly irregular, ramified structures. A prominent example is diffusion-limited aggregation (DLA), which generates fractal clusters characterized by non-integer fractal dimensions ($D_f = 5/3$ in 2D square lattice) and pronounced branching morphology arising from screening effects during growth~\cite{Witten1981}. While DLA has been extensively studied as a growth process, its role as a deposited object in RSA-type models remains largely unexplored.

%DLA cluster has direct experimental realization in different systems, e.g., growth of bacterial colonies of {\it{Bacillus subtilis}} in a self-similar way on a two-dimensional plate with fractal dimension $1.73\pm0.02$~\cite{1990_matsushita_phyA, Tronnolone2018} and electrolytical growth of copper on a point-like cathode from aqueous solution with fractal dimension $2.43\pm0.03$~\cite{Brady1984}. The Hausdorf dimensonality associated with the fractal growth matches with the computer simulation predictions~\cite{Meakin1983PRA} both in two and three dimensions.

%In parallel, previous work has investigated RSA on complex substrates, including fractal and disordered lattices, demonstrating that the underlying geometry of the substrate can significantly influence jamming behavior and scaling properties~\cite{Pasinetti2019}. However, the complementary problem, RSA of fractal objects on regular substrates has received comparatively little attention.

\section{Model \& Algorithm}
We consider an initially empty $L\times L$ square lattice with periodic boundary conditions along both directions. Objects in the form of DLA clusters of a given size $k$ are sequentially deposited at random onto the lattice sites, obeying excluded-volume interactions. The deposition attempts continue until the DLA cluster under consideration can not be accommodated onto any available empty sites. At this stage, the system is said to be in the ``jamming state''. The entire process starting from an empty lattice and reaching to the jamming state, constitute a single run. If $N$ clusters are adsorbed at jamming, the jamming density is defined as $p_j=Nk/L^2$. Depending on whether different DLA cluster shapes are used across different runs, we classify two model variants.

\begin{figure}[t]
    \centering
    \includegraphics[width=0.96\linewidth]{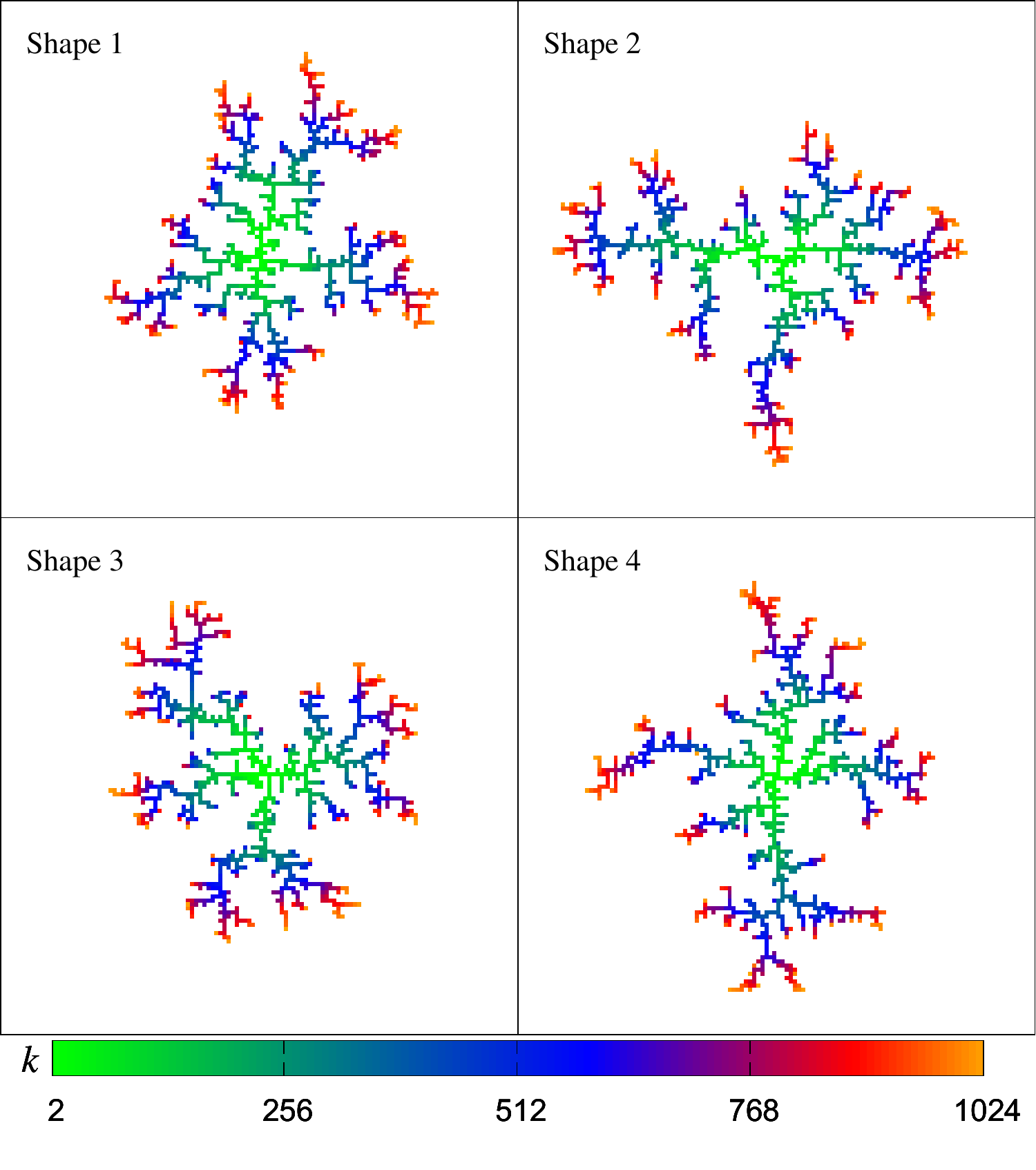}
    \caption{Typical DLA clusters on a square lattice at different stages of its growth process up to size $k=1024$. Four different panels represent independently generated DLA clusters.}
    \label{fig:mono_shapes}
\end{figure}

\emph{Monodisperse fixed-shape:} As in the standard RSA, where a given regular shape is used for deposition, in this variant, a fixed DLA cluster shape of size $k$ is used for all deposition attempts within a single run and across different runs as well. 
%To increase the cluster size from $k$ to $k+m$, an additional $m$ particles are added to the same growing cluster. 
Figure~\ref{fig:mono_shapes} shows four representative DLA cluster shapes used in our study.

\emph{Monodisperse refreshing-shape:} In contrast to the above variant, here, a new cluster shape is generated independently for each run and used throughout the run.

The entire deposition process involves two main computationally time consuming tasks: (i) generating DLA clusters, and (ii) efficiently tracking the deposition process to determine whether the jamming state is reached.

%Random sequential adsorption is a process in which objects of different geometrical shapes and sizes are sequentially deposited on a confined space. Particles interact among themselves only through hard-core or excluded volume interaction. Universality that emerges in jamming and percolation transitions are due to the local screening effect of adsorbed particles. In this paper we study RSA model of DLA clusters on two-dimensional square lattice of size $L\times L$ with periodic boundary along two directions. We explore the dependence of size and shape diversity of DLA-clusters on jamming and percolation transitions. In the following two subsections we describe the algorithm that we use to generate a DLA cluster of particular size and describe the algorithm to reach the jamming state.

\subsection*{Generating DLA clusters}
%The ideal way of generating a DLA cluster is by fixing the initial seed at a central point and releasing particles one by one from a large distance chosen randomly obeying the angular symmetry with respect to the central point. The released particle does random walk until it meets the initial seed and sticks to it. The process is repeated until we get a DLA cluster of desired size. The efficiency of the algorithm largely depends on the distance of the release point from the central point. We set a suitable intermediate value of the distance that make the DLA generation fast enough and at the same time the geometrical properties of the cluster, e.g., $D_f = 5/3$ is kept intact.

We use the DLA model introduced by Witten and Sander for cluster generation~\cite{Witten1981}. The DLA process starts with a single seed particle placed at the center of an empty square lattice. A new particle is then released from a randomly selected position sufficiently far from the seed and performs a random walk until it reaches a site adjacent to the occupied site, where it irreversibly attaches to form a two-particle cluster. Subsequent particles are released in a similar manner far from the growing cluster and undergo the random-walk until they become part of the cluster. This process is repeated until the desired cluster size $k$ is reached.

%The process starts on a separate empty lattice with a single seed particle placed at the center. New particles are released successively from randomly selected positions sufficiently far from the growing cluster and perform random walks until they encounter the cluster. A particle is incorporated into the cluster when it reaches a site adjacent to one of the peripheral sites of the cluster. This process is repeated until the cluster reaches the desired size $k$.

To gain computational efficiency, at each step, we determine the maximum radial extent $r_{\rm max}$ of the growing cluster centered on the seed site and dynamically adjust the radius of the release circle according to $r=r_0+r_{\rm max}$, as in~\cite{Meakin1983}. A new particle is released from a random point on this circle that performs a random walk. If the particle moves beyond the killing radius $r=r_0+r_{\rm max}+\Delta r$, it is discarded and a new particle is released. On the other hand, the proximity of the particle to the existing cluster is checked only when it is within the concentric circle of radius $r_{\rm max}+2$. In our simulations, we set $r_0=100$ and $\Delta r=20$, in units of lattice constant. With this set of parameters, we generate large DLA clusters containing up to $10^5$ particles and verify their fractal dimension using the Box-counting method~\cite{Feder1988}, yielding $D_f=1.67\pm0.02$.

\subsection*{Determining jamming state}\label{Subsec_deposition}

\begin{figure}[t]
\includegraphics[width=0.98\linewidth]{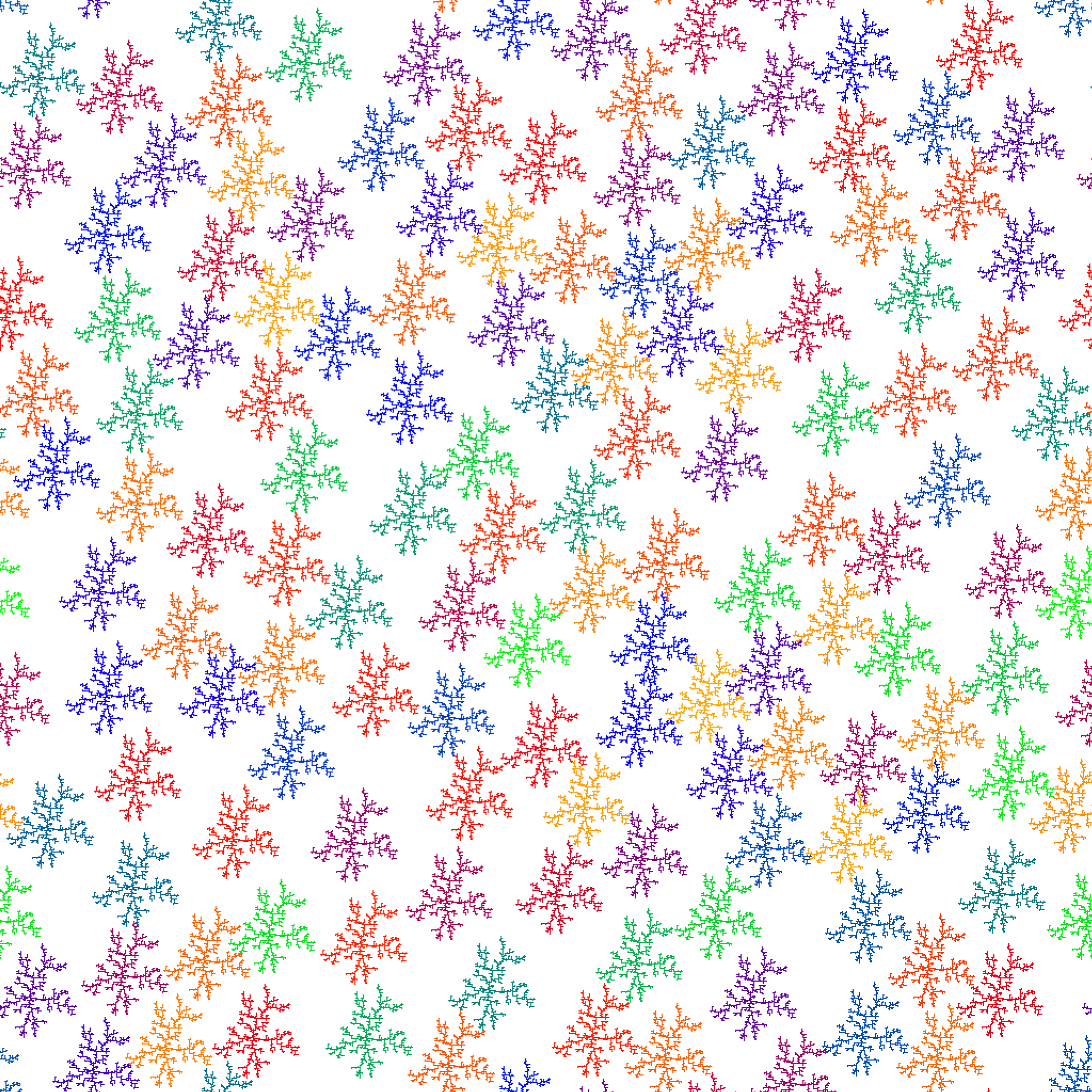}
    \caption{Typical jamming configuration for DLA clusters of size $k=1024$ in the \textit{monodisperse fixed-shape} model using shape-1 on a square lattice of size $L=1024$ with periodic boundary conditions.}
    \label{fig:jamming_state}
\end{figure}

In both variants of the model, we implement a rejection-free algorithm to ensure that an empty site is always selected in every deposition attempt. Initially, a list of all empty sites is prepared, from which a random site is selected. The list is updated after each deposition attempt. If the DLA cluster is successfully deposited, all sites occupied by the cluster are removed from the list; otherwise, only the selected site is removed. Thus, the number of sites in the list decreases iteratively, and the process ends when the list becomes completely empty. This corresponds to the jamming state.

In the most general case, a new DLA cluster would be generated for each deposition attempt in a single run. Such an approach is computationally expensive, since the generation of a DLA cluster is itself time-consuming. A further difficulty arises in defining the jamming state, because the configurational space of number of possible DLA shapes becomes extremely large for a sufficiently large DLA cluster. One possibility is to terminate the deposition process after a certain number of consecutive attempts fail to place a newly generated cluster on the available empty sites. 

However, our objective here is different: we aim to systematically investigate how the degree of DLA shape diversity impacts the properties of the jamming state. For this reason, we consider different variants of our deposition model. Notably, the \textit{monodisperse fixed-shape} and \textit{monodisperse refreshing-shape} protocols exhibit substantially different scaling behavior of the jamming density, as discussed in Sec.~\ref{sec:scaling}. Furthermore, we consider a \textit{polydisperse} variant in which each deposition run uses a pool consisting of $S$ independently generated DLA cluster shapes, with $S$ controlling the degree of shape diversity (see details in Sec.~\ref{sec:poly}).

%For each cluster size, we use a fixed cluster shape, which is shown in Fig.~\ref{fig:mono_shapes} for some of the cluster sizes used in our study. The deposition process works as follows: We pick a lattice site randomly and attempt to deposit the cluster. If deposition is successful, we move forward on to the next deposition attempt on a new random site. If deposition fails, we repeat the deposition process on a new random site. We repeat this process until there are no more available sites where the cluster can be deposited, at which point we consider the system to be in a jammed state. By repeating the deposition process multiple times with different random initializations using the same cluster shape, we can calculate the average jamming density and other properties of the system.

The simulations were performed on the HPC system with Intel Xeon Platinum 8380 processors (2.30 GHz) and 1 TB of memory per node. We employed OpenMP parallelization utilizing 160 threads available per node. To give an idea about the computational complexity, we report the computational wall-time in hours (h) required to generate 1000 jamming configurations for $k=256$. The monodisperse \textit{fixed-shape} and \textit{refreshing-shape} variants on $L=2048$ required approximately 0.12 h time. The polydisperse fixed-pool variant on $L=1024$, the corresponding times are approximately 0.33 h and 303 h for pool sizes $S=10^3$ and $10^6$, respectively. Finally, \textit{polydisperse refreshing-pool} variant on $L=1024$ required 0.81 h time for $S=10^3$.

\section{Results}
\subsection{Jamming density}
\begin{figure}[t]
    \centering
    \includegraphics[width=1.0\linewidth]{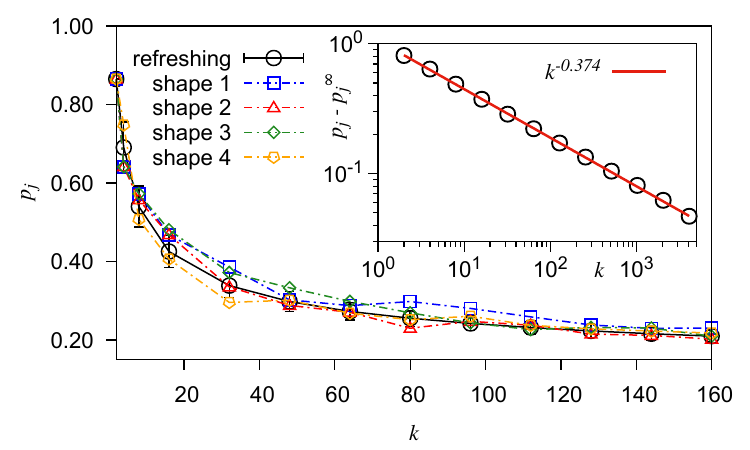}
    \caption{Jamming density $p_j$ as a function of DLA cluster size $k$ for the four cluster shapes displayed in Fig.\ \ref{fig:mono_shapes} in the \textit{monodisperse fixed-shape} variant and for the \textit{monodisperse refreshing-shape} variant. Data are based on lattice size $L=2048$ and averaged over $10^5$ configurations for $k\geqslant32$ and at least $10^4$ configurations for $k<32$. (Inset) Log-log plot of $p_j-p_j^\infty$ against $k$ for the \textit{ refreshing-shape} variant.}
    \label{fig:mono_AvPool_vs_FixPool_jd}
\end{figure}

We first estimate the average jamming density $p_j$ for the \textit{monodisperse fixed-shape} case for the four different cluster shapes shown in Fig.~\ref{fig:mono_shapes}. A snapshot of the jamming configuration using shape 1 with cluster size $k=1024$ on $L=1024$ is shown in Fig.~\ref{fig:jamming_state}. Averaging over many such independent jamming configurations, we obtain the value of $p_j$ for a given shape and size. To examine the effect of increasing the size $k$ of the cluster on $p_j$, we consider the same growing DLA cluster at different stages of its evolution, successively adding particles until the desired size $k$ is reached. In the main panel of Fig.\ \ref{fig:mono_AvPool_vs_FixPool_jd}, the results for $2 \leqslant k \leqslant 160$ are shown for the four different DLA cluster shapes. As the selected shapes are geometrically distinct, their impact on the jamming density $p_j$ is evident. It is important to note that due to growth instability and nonlocal screening~\cite{Sander2000}, different branches of a DLA cluster develop unevenly during its growth process. On rare occasions, a diffusing particle penetrates deep into the DLA cluster, traversing its branched structure, and ultimately attaches within the interior. Such events contribute to the cluster size without altering its spatial extent. As a result of this interplay between size and spatial extent, the effective excluded area of an absorbed DLA cluster increases monotonically with $k$, albeit in a stochastic manner. The effect of this is expected to be reflected in the jamming density. For example, $p_j$ is higher for shape-3 ($p_j= 0.3333$) than for shape-1 ($p_j=0.3025$) at $k=48$, whereas the opposite trend is observed at $k=80$, with $p_j=0.2697$ and $0.2983$ for shape-3 and shape-1, respectively (see also Fig.\ \ref{fig:jd_distr-Av_vs_Fix}). These results are statistically accurate, based on $10^5$ independent realizations for each shape on a $L=2048$ lattice. The jamming density $p_j(L)$ exhibits negligible finite-size effects (not shown) when $L/R_g(k)$ remains sufficiently larger than unity, with $R_g$ being the radius of gyration. This is ensured in all our simulations, in particular, we have $\bm\min(L/R_g(k))=32$. For cluster sizes up to $k=4096$, the average radius of gyration follows,
\begin{equation}
\langle R_g(k) \rangle = 0.496 k^{0.582}.
\end{equation}
Since $R_g(k)\sim k^{1/D_f}$, this scaling yields a fractal dimension $D_f=1/0.582\simeq1.718$ over the range $k\leqslant 4096$.

The number of distinct DLA shapes, i.e., the configurational space, increases markedly with increasing $k$. Therefore, in the \textit{refreshing-shape} case, DLA clusters generated in different runs are expected to be distinct with high probability for a sufficiently large value of $k$. Figure~\ref{fig:mono_AvPool_vs_FixPool_jd} shows that the $p_j(k)$ as a function of $k$ obtained from the \textit{refreshing-shape} follows a trend similar to that of the \textit{fixed-shape} case; averaging over $10^5$ independently generated shapes only smooths the dependence.

The jamming density monotonically decreases with increasing $k$ for \textit{refreshing-shape} case. A least-squares fit to the data on a logarithmic scale over the range $2 \leqslant k \leqslant 4096$ shows that the decay is well described by a power-law 
\begin{equation}
p_j(k)-p_j^\infty \sim k^{-\alpha}, 
\label{power-law}
\end{equation}
with $\alpha=0.374(2)$ and $p_j^\infty = 0.051(2)$. This is shown in the inset of Fig.~\ref{fig:mono_AvPool_vs_FixPool_jd}. The extrapolated value of $p_j^\infty$ can be interpreted as the jamming density in the limit of an infinitely large DLA cluster. Since $L/R_g(k)\gg1$ is maintained even in the limit $k\to\infty$, a finite value of $p_j^\infty$ is expected to persist. As used in the case of SAW~\cite{Ramirez2023}, we also tried to fit our data using a quadratic polynomial $p_j(k)=A+B/k+C/k^2$, with $A, B$ and $C$ as parameters. The polynomial fit represents the data well for $k\lesssim12$, but the deviation becomes increasingly pronounced at larger $k$.
%\textcolor{blue}{The quadratic polynomial fit works for small $k$, but it deviates for large $k$. A Power-law fit works better to represent the behavior.} 
%To confirm this we calculate the fractal dimension $d_{fj}$ of the jammed state and we found that $d_{fj}=d_f \approx 1.67$, which in turn gives us $p_j \sim k^{d_{fj}-2}$.} %\SK{How can we argue that the jammed state will have same fractal dimension of a single DLA cluster? For SAW is 4/3, so can we see $p_j(k) \sim k^{-2/3}$?}

%following a $\lambda^{-1/3}$ dependence.

%Our simulations of the monodisperse case reveal interesting trends when comparing the fixed shape and shape averaged scenarios. We performed $10^6$ configurations for the fixed shape case and averaged over $10^6$ different shapes for the shape averaged case. The results show that the data becomes smoother in the shape averaged case, indicating a more consistent behavior across different realizations. However, the standard deviation of the jamming density increases in the shape averaged case, likely due to the fact that each shape has a different jamming density, leading to a broader distribution of values. The plot of jamming density as a function of DLA size $\lambda$ shows a power-law relation, with the data closely 
\begin{figure}[t]
    \centering
    \includegraphics[width=0.98\linewidth]{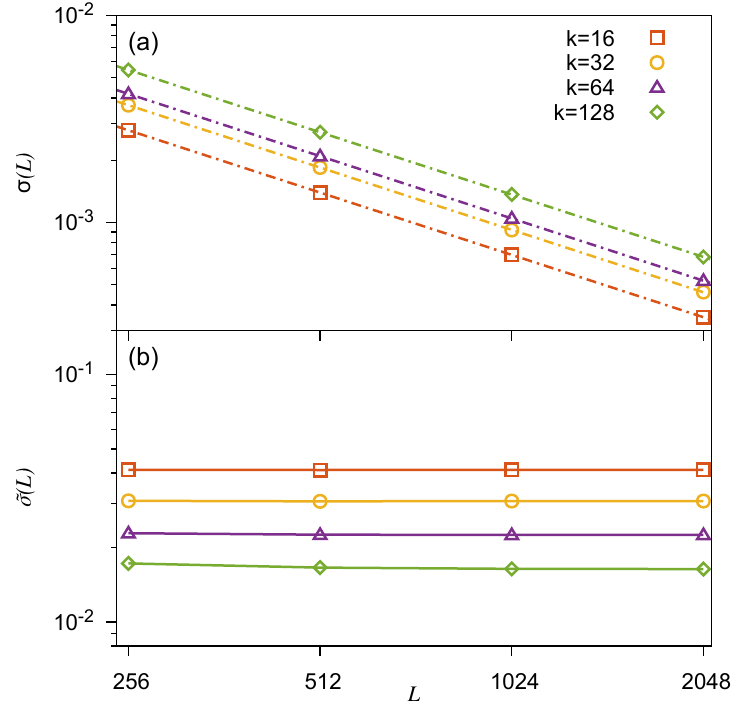}
    \caption{Scaling behavior of the fluctuation of jamming density. (a) Log-log plot of the standard deviation $\sigma(L)$ against system size $L$ for the \textit{monodisperse fixed-shape} variant. The dotted line shows a power-law fit $L^{-1.0001(5)}$. (b) Corresponding results $\tilde{\sigma}(L)$ for the \textit{refreshing-shape} variant, showing no apparent dependence on $L$. Results are obtained by averaging over $10^5$ realizations.}
    \label{fig:mono_jamming_exp}
\end{figure}
\subsection{Fluctuation of the jamming density}\label{sec:scaling}
We define the fluctuation of the jamming density as
\begin{equation}
\sigma(L)=(\langle p_j^2 \rangle - \langle p_j \rangle^2 )^{1/2}, 
\end{equation}
where $\langle .. \rangle$ represents the ensemble average. Typically, $\sigma(L)$ scales as $\sigma(L) \sim L^{-1/\nu_j}$, with $\nu_j=2/D$ for a D-dimensional substrate~\cite{Pasinetti2019}, yielding $\nu_j=1$ for $D=2$. Many variants of the RSA model, including those with randomly distributed defects~\cite{Ramirez_defect_2019}, exhibit the same exponent $\nu_j=1$ in 2D, indicating the universality of the jamming transition. This universal scaling is found to break down only in the presence of long-range spatially correlated defects~\cite{Kundu2021}. To check whether universal scaling law holds for this DLA cluster deposition models, in Fig.~\ref{fig:mono_jamming_exp}(a), we plot $\sigma(L)$ as a function of $L$ on a log-log scale for different system sizes $L$ ranging from 256 to 2048, and determine the exponent value by measuring the slope of the curves.

For the \textit{monodisperse fixed-shape} variant, the estimated exponent $\nu_j=1.0001(5)$ agrees well with its standard value in 2D, as shown in Fig.~\ref{fig:mono_jamming_exp}. In contrast, the fluctuation in jamming density for the \textit{refreshing-shape} variant exhibits distinctly different behavior, as shown in Fig.~\ref{fig:mono_jamming_exp}(b). In particular, $\sigma(L)$ shows no apparent dependence on $L$; instead, it takes a constant value that depends on the DLA cluster size $k$. We explain this behavior in the following way.
\begin{figure}
    \centering
    \includegraphics[width=0.98\linewidth]{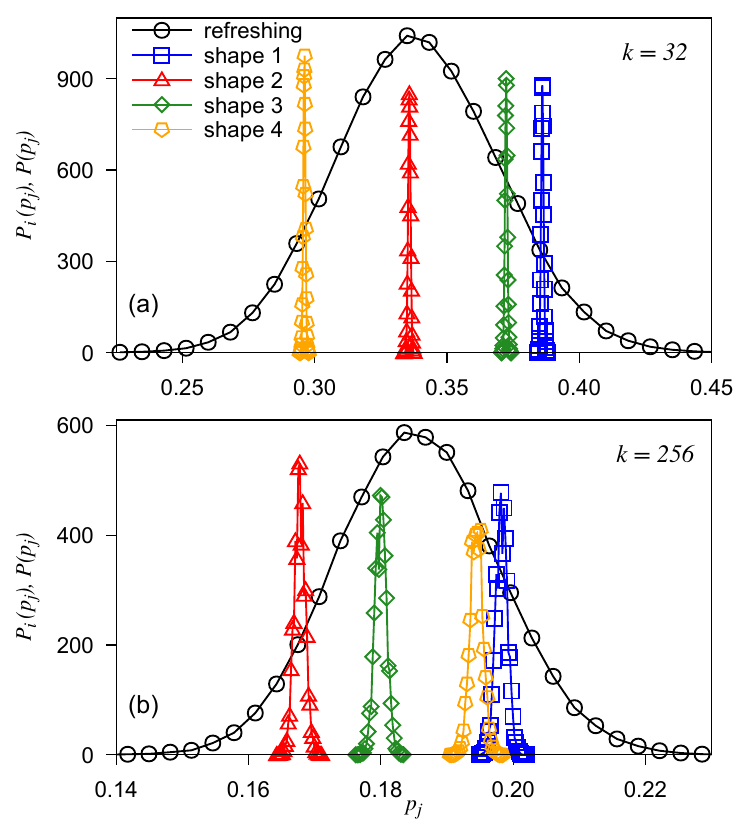}
    \caption{Probability density functions of the jamming density, $P_i(p_j)$ for the \textit{monodisperse fixed-shape} and $P(p_j)$ for the \textit{monodisperse refreshing-shape} variants for (a) $k=32$ and (b) $k=256$. The distribution width increases with $k$ for the \textit{fixed-shape} variant, whereas it decreases for the \textit{refreshing-shape} variant. Results are averaged over $10^5$ realizations on $L=2048$. For better visualization, $P(p_j)$ curves are scaled up by factors of $80$ and $18$ in panels (a) and (b), respectively.}
    \label{fig:jd_distr-Av_vs_Fix}
\end{figure}

Since the DLA cluster shape is updated after each realization of the jamming configuration in the refreshing-shape variant, the jamming density and its fluctuation can be interpreted as the result of an ensemble of jamming configurations for distinct fixed-shapes. More elaborately, in the refreshing-shape variant, one can, in principle, generate an infinite number of realizations. Now, for an intermediate value of $k$, the number of distinct shapes is finite, and thus each shape necessarily appears in multiple realizations. The ensemble of refreshing-shape runs can therefore be grouped into disjoint subsets, each corresponding to realizations associated with a given fixed shape $i$. Based on this, we construct a probabilistic explanation of the observed behavior of $\sigma(L)$.

Let $n$ be the number of distinct possible shapes for a given size of the DLA cluster $k$ and $P_i(p_j)$ be the probability density function of the jamming density $p_j$ corresponding to a fixed shape $i$, with mean $\mu_i=\langle p_j \rangle$ and standard deviation $\sigma_i$.
The probability density of $p_j$ for the refreshing-shape variant is thus given by
\begin{equation}
    P(p_j)=\frac{1}{n}\sum_{i=1}^nP_i(p_j).
    \label{eq:prob_pj}
\end{equation}
Here, we assume that all distinct shapes occur with equal probability, which is expected for sufficiently large DLA clusters but may be violated for small clusters.
In Figs.~\ref{fig:jd_distr-Av_vs_Fix}(a) and (b), we show the probability density function $P_i(p_j)$ for the \textit{monodisperse fixed-shape} variant for the four different DLA cluster shapes, together with the corresponding distribution $P(p_j)$ for the \textit{refreshing-shape} variant. As expected, $P(p_j)$ spans the range of jamming densities associated with the individual cluster shapes for a given $k$. The mean jamming densities $\mu_i$ show no appreciable finite-size dependence, although the distribution width for a fixed shape scales as $\sigma_i(L)\sim 1/L$. Thus, as $L$ increases, the individual shape distributions become narrower around their respective mean $\mu_i$. Consequently, the spread of the combined distribution for the \textit{refreshing-shape} variant remains finite and shows no appreciable finite-size dependence.

\begin{figure}[t]
    \centering
    \includegraphics[width=0.98\linewidth]{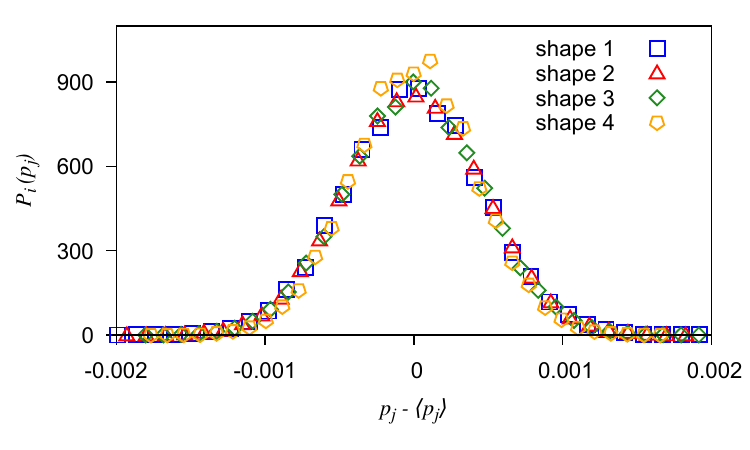}
    \caption{For fixed-shape and $k=32$, the probability density distribution $P_i(p_j)$ data as in Fig.\ \ref{fig:jd_distr-Av_vs_Fix}(a) shows excellent collapse when plotted as a function of $(p_j-\langle p_j\rangle)$.}
    \label{fig:sigma_i}
\end{figure}

A more theoretical explanation can be given as follows. We assume that $P_i(p_j)$ follows a Gaussian distribution with an identical standard deviation $\sigma_i=\sigma$, as verified for the four fixed shapes (see Fig.~\ref{fig:sigma_i}). %Probability distribution for individual shape shows excellent collapse when plotted as a function of $(p_j-\langle p_j\rangle)$. 
Accordingly, the variance $\tilde{\sigma}^2$ of the distribution $P(p_j)$ can be calculated using Eq.\ \ref{eq:prob_pj} as
\begin{equation}
\label{sigma_refresh}
    \tilde{\sigma}^2=\sigma^2+\bigg(\frac{1}{n}\sum_{i=1}^n\mu_i^2-\mu^2\bigg),
\end{equation}
where $\mu=1/n\sum_{i=1}^n\mu_i$. Since $\sigma(L) \sim 1/L$ for the fixed-shape case, the first term in Eq.\ \ref{sigma_refresh} is $\mathcal{O}(1/L^2)$. On the other hand, $\mu_i\sim\mathcal{O}(1)$. This implies that, for intermediate values of $k$, the leading-order behavior of $\tilde{\sigma}$ is determined by the second term in Eq.~\ref{sigma_refresh}, which is a $L$-independent constant value. Importantly, $\{ \mu_i\}$ values depends on $k$ and thus, the constant is $k$-dependent.

\begin{figure}[t]
    \centering
     \includegraphics[width=0.98\linewidth]{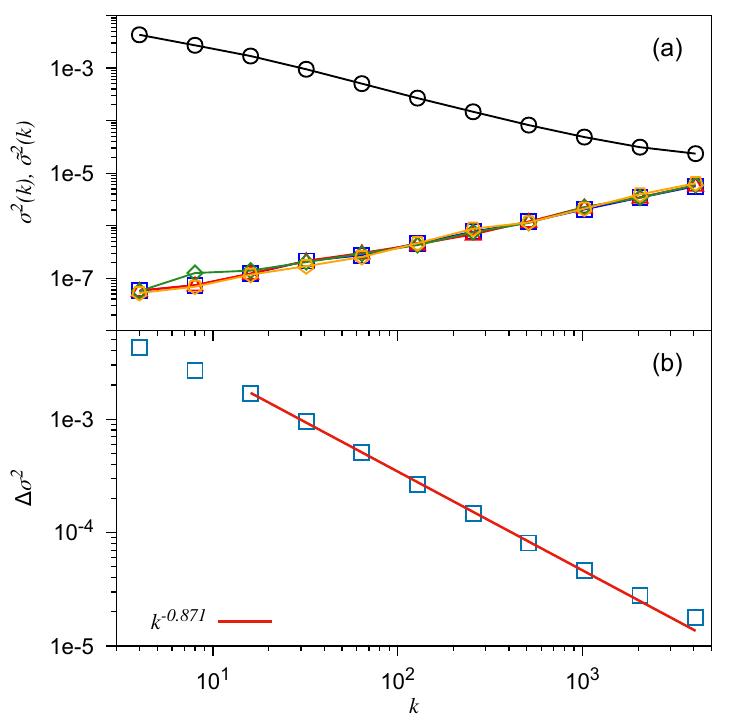}
    \caption{(a) Variance of the jamming density against cluster size $k$ for four distinct shapes in the \textit{fixed-shape} variant and for the \textit{refreshing-shape} variant. The same symbols and statistics as in Fig.~\ref{fig:jd_distr-Av_vs_Fix} are used to represent the data. (b) The difference in variance between the \textit{refreshing-shape} and shape-1 in the \textit{fixed-shape} variants vanishes as a power-law, suggesting convergence of the two variants in the limit $k\to\infty$.}
    \label{fig:Mono_SD_vs_k}
\end{figure}

Now we investigate how the variance of the jamming density distribution depends on $k$ for both variants. Interestingly, for a fixed shape, $\sigma(k)$ increases with $k$, whereas the corresponding width $\tilde{\sigma}(k)$ for the \textit{refreshing-shape} variant decreases, as shown in Fig.~\ref{fig:jd_distr-Av_vs_Fix} and more explicitly in Fig.~\ref{fig:Mono_SD_vs_k}(a). This plot further suggests that the difference ${\Delta\sigma}^2 = \tilde{\sigma}^2(k) - \sigma^2(k)$ decreases with increasing $k$ and tends to approach zero in the large-$k$ limit. To verify this, we plot ${\Delta\sigma}^2(k)$ as a function of $k$ on a log-log scale in Fig.~\ref{fig:Mono_SD_vs_k}(b). The data is well described by a power-law ${\Delta\sigma}^2(k) \sim k^{-0.871(8)}$. This implies that, in the thermodynamic limit $k\to\infty$ and $L\to\infty$, while maintaining $L/R_g(k)\gg1$, the \textit{fixed-shape} variant converges to the \textit{refreshing-shape} variant. We argue that this behavior follows from the statistical self-similarity of DLA clusters in the large-$k$ limit. Consequently, the mean jamming densities for the fixed-shape and refreshing-shape variants are expected to become identical, $\mu_i=\mu$ for all $i$, in the limiting case, which implies $\tilde{\sigma}=\sigma$. The case $k=2$ is special, for which $\tilde{\sigma}=\sigma$ holds trivially.

\section{Effect of polydispersity on jamming density}{\label{sec:poly}}
In this section, we generalize our model by considering different DLA cluster shapes to be deposited in a single run. The number of distinct shapes used in a run determines the degree of polydispersity. To systematically investigate its effect on the jamming properties of the system, we introduce a ``cluster pool'' of size $S$, containing $S$ independently generated DLA clusters of size $k$. For small $k$ and large $S$, a particular shape may appear multiple times in the pool. Such repetitions become increasingly unlikely as $k$ increases and are negligible for sufficiently large $k$. However, statistical self-similarity of clusters becomes relevant at that length scales.

The deposition process proceeds as follows: for each step, a cluster is selected randomly from the pool and an empty lattice site is chosen at random for deposition. If the deposition attempt is successful, we begin the next trial attempt by drawing a cluster at random from the same pool. If the deposition attempt fails, we pick another site and try depositing the same cluster. If it fails again, we pick another site, and so on. If the cluster fails to deposit on all available sites, it is removed from the pool.
%Otherwise, deposition of the same cluster is attempted sequentially on other empty positions at random until either the cluster is successfully deposited or all available sites are exhausted. In the latter case, the pool size is decreased by one by removing the cluster from the pool. 
For the next deposition attempt, we randomly select a cluster from the available clusters in the pool. The system is defined to be in jamming state when the cluster pool becomes completely empty. This process generates a single jamming configuration. As in the previous cases, we employ the rejection-free algorithm for site selection. Note that the model contains additional sources of randomness beyond the random selection of lattice sites that enters through the shape diversity of DLA clusters.

As before, two variants of the algorithm are considered. In the \textit{polydisperse fixed-pool} case, a pool consisting of a set of DLA cluster shapes is used across runs. In the \textit{refreshing pool} case, the pool is reinitialized in every run by randomly generating a new set of DLA clusters.

\begin{figure}
    \centering
    \includegraphics[width=0.98\linewidth]{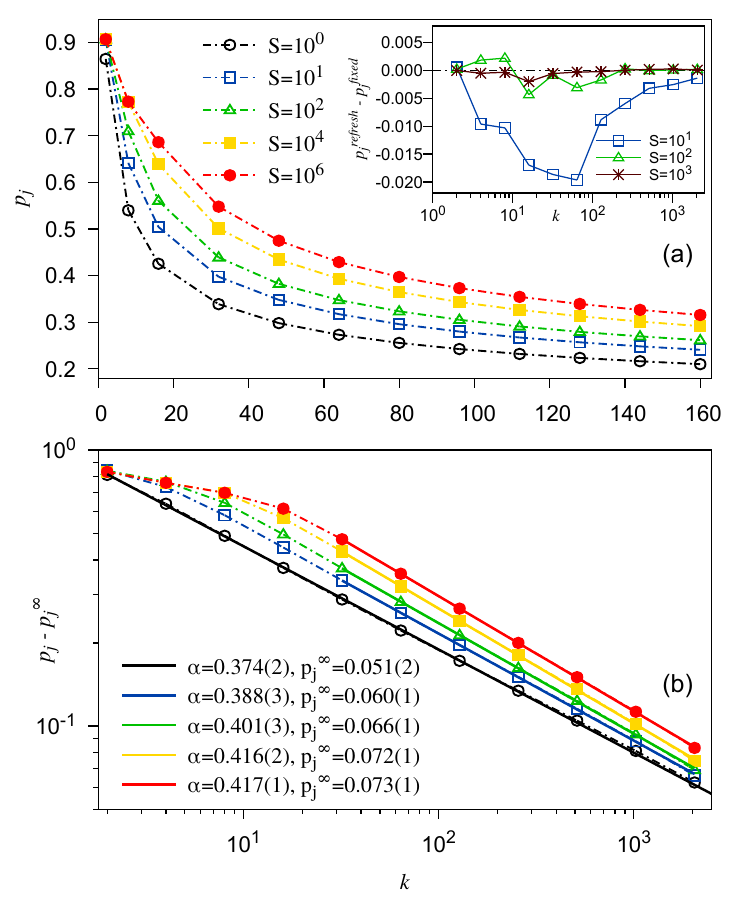}
    \caption{(a) Jamming density $p_j$ versus cluster size $k$ for different pool sizes $S$ in the \textit{polydisperse refreshing-pool} ($S=10$ and $10^2$, open symbols) and \textit{polydisperse fixed-pool} ($S=10^4$ and $10^6$, filled symbols) variants on $L=1024$, together with the \textit{monodisperse refreshing-shape} case ($S=1$). The inset shows the difference in $p_j$ between the \textit{refreshing-pool} and a particular \textit{fixed-pool}, which decreases with increasing $S$. Results are obtained by averaging over (at least) $5\times10^4$ for $S\leqslant10^4$ and $500$ for $S=10^6$ independent realizations. (b) Power-law fits (solid lines) to the same data shown in (a) using Eq.\ \ref{power-law}. The best-fit parameter values are listed in the plot.}
    \label{fig:AvPool_Mono_vs_Poly200_L1024}
\end{figure}

In Fig.~\ref{fig:AvPool_Mono_vs_Poly200_L1024}(a), we show the variation of the jamming density as a function of $k$ with $S=10, 10^2$ for the refreshing-pool case and $S=10^4, 10^6$, for the fixed-pool case, together with the monodisperse refreshing case ($S=1$). The curves exhibit a systematic upward shift with increasing $S$ over the entire range of $k$. This implies that the jamming density increases with increasing the degree of polydispersity. Therefore, polydispersity promotes denser packing of the clusters.

\begin{figure}[t]
    %\centering
    \includegraphics[width=0.98\linewidth]{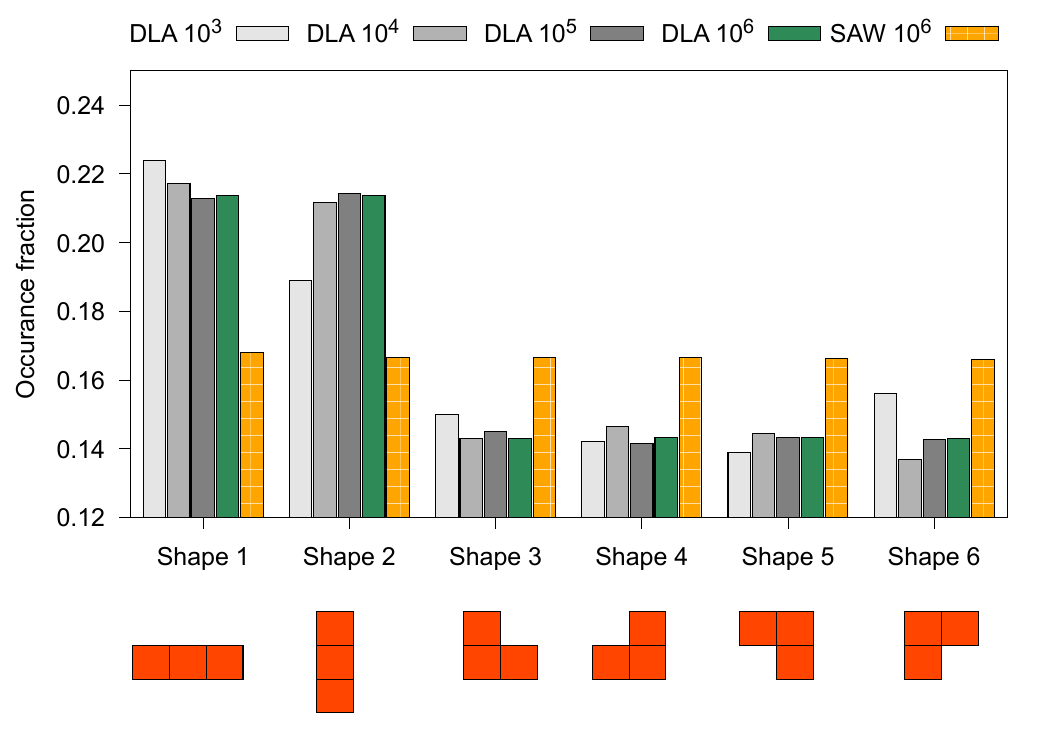}
    \caption{Cluster shape statistics for DLA and SAW for $k=3$. The relative fractions of all possible DLA shapes are shown for pool sizes ranging from $S=10^3$ to $10^6$. As $S$ increases, these fractions converge toward their nonuniform asymptotic occurrence probabilities. In contrast, the six SAW shapes have equal occurrence probabilities of $1/6$, illustrating how the sampling of shapes is process dependent.}
    \label{fig:shape_dla_vs_saw}
\end{figure}

In addition, the plot in the main panel suggests that the difference between consecutive $p_j$ curves decreases with increasing $S$, indicating that the curves approach a limiting curve corresponding to an infinitely large pool size. Qualitatively, this behavior can be understood as follows. Consider an intermediate value of $k$, for which the number of distinct DLA shapes is moderately large. The underlying DLA process generates different shapes with nonuniform occurrence probabilities. For example, a linear shape is less likely to occur than branched configurations. Now, for small $S$, the pool contains only a limited subset of the possible DLA shapes. Therefore, the relative fractions of the sampled shapes may differ substantially from their occurrence probabilities. As $S$ increases, the pool gradually incorporates a larger fraction of the distinct possible shapes, and as a result, their relative fractions approach the corresponding occurrence probabilities (see Fig.\ \ref{fig:shape_dla_vs_saw} for an example). This qualitatively explains the decrease in the difference between consecutive $p_j$ curves with increasing $S$. Further, following this argument, in the large-$S$ limit, the distinction between both variants should disappear and both should yield nearly the same $p_j$ values. This is illustrated in the inset of Fig.~\ref{fig:AvPool_Mono_vs_Poly200_L1024}(a), where the negligible difference is particularly visible for $S=10^3$.

Furthermore, Fig.\ \ref{fig:AvPool_Mono_vs_Poly200_L1024}(a) hints at the convergence of all the curves, including the monodisperse case, in the thermodynamic limit of $k \to \infty$. This convergence again follows from the statistical self-similarity of DLA clusters in the large-$k$ limit.

Plotting the same data as in Fig.~\ref{fig:AvPool_Mono_vs_Poly200_L1024}(a) on a log-log scale reveals the existence of a lower cutoff above which the data exhibit signatures of power-law behavior. The cutoff becomes increasingly pronounced with increasing pool size $S$. Therefore, here as well, we perform a power-law fit to the data using Eq.~\ref{power-law}, but beyond the visually determined cutoff, as shown in Fig.~\ref{fig:AvPool_Mono_vs_Poly200_L1024}(b). The exponent $\alpha$ varies from $0.374(2)$ ($S=1$) to $0.417(1)$ ($S=10^6$).

\begin{figure}
    \centering
    \includegraphics[width=0.98\linewidth]{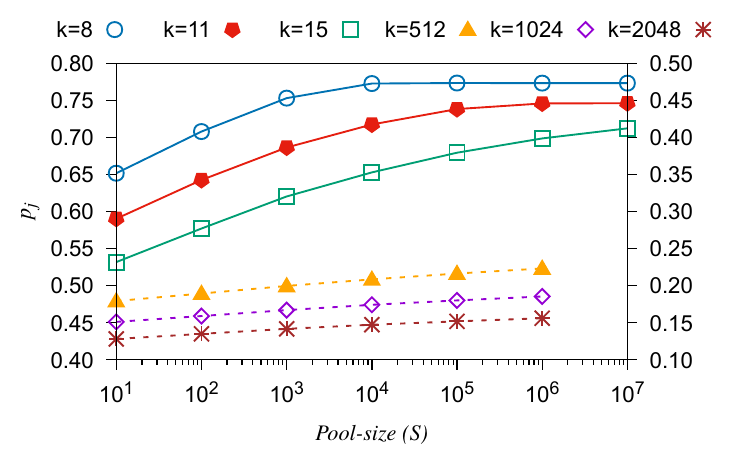}
    \caption{Dependence of jamming density $p_j$ on the pool size $S$ for the \textit{polydisperse fixed-pool} variant on $L=1024$ for various $k$. For $k\leqslant11$ saturation of $p_j$ with increasing $S$ is apparent, whereas for large-$k$, only a slow variation of $p_j$ is observed with $S$, possibly because of the self-similarity of DLA clusters. For $k\geqslant512$ (dashed lines), the $p_j$ axis is shown on the right. Number of realizations are same as in Fig.\ \ref{fig:AvPool_Mono_vs_Poly200_L1024}.}
    \label{fig:singlek_pj_pooldep}
\end{figure}

Figure \ref{fig:singlek_pj_pooldep} demonstrates the existence of a threshold pool size $S_c(k)$, beyond which the jamming density $p_j$ saturates. For small values of $k$, this saturation of $p_j$ is particularly noticeable. The value of $S_c(k)$ increases with $k$. The lower cutoff observed in Fig.~\ref{fig:AvPool_Mono_vs_Poly200_L1024} may be related to the existence of $S_c(k)$. In particular, for $k=8$ and 11, $p_j$ saturates at $S_c(k)=10^4$ and $10^6$, respectively, and these $k$ values are approximately the lower cutoff values for $S=10^4$ and $10^6$ in Fig.~\ref{fig:AvPool_Mono_vs_Poly200_L1024}(b). 

At large-$k$, because of the large configurational space, and the statistically self-similar fractal nature of the clusters, $p_j$ varies only weakly with $S$. In particular, we find $p_j(S=10^6)-p_j(S=10) \approx 0.028$ for $k=2048$. Whether a value of $S_c$ exists for such large fractal clusters remains unclear to us and require further investigation, which are beyond the scope of our computational resources.

Finally, we remark that our polydisperse variant for $k=3$ with $S \to \infty$, does not exactly reduce to the RSA of SAW chains on a square lattice~\cite{Ramirez2023}. The difference lies in the occurrence probabilities of the distinct shapes. In SAW, each shape occurs with equal probability ($1/6$), whereas in DLA the corresponding probabilities are nonuniform, as presented in Fig.\ \ref{fig:shape_dla_vs_saw}. For example, based on $10^6$ randomly generated DLA configurations, we estimate the probability of each linear shape to be approximately $0.2125$ in DLA. This difference is reflected in the corresponding jamming densities. Specifically, we obtain $p_j=0.86094(1)$ for DLA, compared to $p_j=0.85829(2)$~\cite{Ramirez2023} for SAW. As $k$ increases, branched configurations arise in DLA, and our model becomes increasingly different from the RSA of SAW chains. A few $p_j$ values are listed in Table \ref{tab:comp_DLA_SAW} for comparison.

\begin{table}
    %\centering
    \begin{ruledtabular}
    \begin{tabular}{c c c}
        $k$    & DLA    &   SAW\\
        \hline
        %\hline
         3 & 0.86094(1) & 0.85829(2)\\
        %\hline
         4 & 0.83399(1) & 0.83705(2)\\
        %\hline
         9 & 0.76296(1) & 0.76590(9)\\
       %\hline
        11 & 0.74566(1) & 0.75006(5)\\
        %\hline
        15 & 0.69848(1) & 0.73012(6)\\
    \end{tabular}
    \end{ruledtabular}
    \caption{Comparison of $p_j$  values for DLA and SAW deposition on a square lattice. DLA results are obtained using the \textit{polydisperse fixed-pool} variant for $L=1024$ with $S=10^6$ and are averaged over at least $5000$ independent realizations, whereas those for SAW chains are taken from Ref.~\cite{Ramirez2023}.
    } 
    \label{tab:comp_DLA_SAW}
\end{table}

The scaling behavior of the fluctuation of the jamming density, $\sigma(L)$, has also been investigated for both polydisperse \textit{fixed-pool} and \textit{refreshing-pool} variants. Just as a single DLA shape yields a unique jamming density, a fixed pool of DLA shapes here corresponds to a unique value of $p_j$. As expected, $\sigma(L) \sim 1/L$ is observed in this case. In contrast, for the refreshing-pool, the scaling behavior depends on whether the pool size $S$ is large enough to accurately reproduce the occurrence probabilities of the shapes. When this condition is satisfied, the standard scaling of $\sigma(L) \sim 1/L$ is recovered (not shown). Otherwise, the standard behavior breaks down, as observed in the monodisperse refreshing-shape case.

\section{Conclusion}
In summary, we have investigated the RSA of ramified DLA clusters on a square lattice using massively parallel Monte Carlo simulations. Depending on the the degree of DLA cluster shape diversity, we have considered four model variants and characterized jamming properties over a broad range of cluster sizes $2 \leqslant k \leqslant 4096$.
%providing access to a wide range of scales for the study of RSA of fractal objects.

We first studied the \textit{monodisperse fixed-shape} and \textit{monodisperse refreshing-shape} variants to distinguish the effects of using a fixed DLA shape from those of independently generated DLA shapes across runs. We find that the jamming density decreases monotonically with cluster size $k$ as $p_j(k)-p_j^\infty\sim k^{-\alpha}$, with $\alpha=0.374(2)$ and $p_j^\infty=0.051(2)$. Most strikingly, we observe that the fluctuation of jamming density scales as $\sigma(L)\sim1/L$ for the fixed-shape variant, whereas it is independent of $L$ for the refreshing-shape variant. Nevertheless, in the limit $k\to\infty$, the two model variants are expected to converge due to the statistical self-similarity of DLA clusters.

We further introduced polydisperse variants by considering a cluster pool of size $S$ to control shape diversity within and across deposition runs. The jamming density increases with $S$, indicating that polydispersity promotes denser packing. The jamming density again follows the power-law dependence on $k$, but the exponent $\alpha$ varies systematically with the pool size, increasing from $0.374(2)$ for $S=1$ to $0.417(1)$ for $S=10^6$.

Overall, our results highlight the role of shape diversity in determining the jamming properties of fractal objects. While the effect of shape diversity is pronounced at intermediate $k$, statistical self-similarity progressively reduces the differences among the four model variants at large $k$. These findings may provide insights into the design of nanostructured materials with tunable properties utilizing the deposition of fractal clusters~\cite{Fusco2021}.

As a future problem, it would be interesting to study transport properties through the disordered media generated by the deposition of fractal objects at jamming. 
%Diffusion through hyperuniform and porous media has applications in understanding transport dynamics having real world applications~\cite{ncm8-3fmh, PhysRevE_2021_porous_media}. It would be interesting to study the diffusion properties through the occupied sites or vacant sites of jamming configuration formed by the DLA clusters. Specifically, it is worth investigating whether there would be any impression of power-law exponents in the diffusion coefficients.

\section{Data and codes Availability}
Data and codes supporting the study can be made available upon request to the authors.

\section*{Acknowledgment}
The authors gratefully acknowledge the computational and data resources provided through the joint high-performance data analytics (HPDA) project “terrabyte” of the German Aerospace Center (DLR) and the Leibniz Supercomputing Center (LRZ). DM was supported by the European Union - NextGenerationEU under the Italian Ministry of University and Research (MUR) projects PRIN2022:2022NRBLPT-CUP:E53D23001790006.
\balance
%\bibliography{ref}

\begin{thebibliography}{63}%
\makeatletter
\providecommand \@ifxundefined [1]{%
 \@ifx{#1\undefined}
}%
\providecommand \@ifnum [1]{%
 \ifnum #1\expandafter \@firstoftwo
 \else \expandafter \@secondoftwo
 \fi
}%
\providecommand \@ifx [1]{%
 \ifx #1\expandafter \@firstoftwo
 \else \expandafter \@secondoftwo
 \fi
}%
\providecommand \natexlab [1]{#1}%
\providecommand \enquote  [1]{``#1''}%
\providecommand \bibnamefont  [1]{#1}%
\providecommand \bibfnamefont [1]{#1}%
\providecommand \citenamefont [1]{#1}%
\providecommand \href@noop [0]{\@secondoftwo}%
\providecommand \href [0]{\begingroup \@sanitize@url \@href}%
\providecommand \@href[1]{\@@startlink{#1}\@@href}%
\providecommand \@@href[1]{\endgroup#1\@@endlink}%
\providecommand \@sanitize@url [0]{\catcode `\\12\catcode `\$12\catcode
  `\&12\catcode `\#12\catcode `\^12\catcode `\_12\catcode `\%12\relax}%
\providecommand \@@startlink[1]{}%
\providecommand \@@endlink[0]{}%
\providecommand \url  [0]{\begingroup\@sanitize@url \@url }%
\providecommand \@url [1]{\endgroup\@href {#1}{\urlprefix }}%
\providecommand \urlprefix  [0]{URL }%
\providecommand \Eprint [0]{\href }%
\providecommand \doibase [0]{https://doi.org/}%
\providecommand \selectlanguage [0]{\@gobble}%
\providecommand \bibinfo  [0]{\@secondoftwo}%
\providecommand \bibfield  [0]{\@secondoftwo}%
\providecommand \translation [1]{[#1]}%
\providecommand \BibitemOpen [0]{}%
\providecommand \bibitemStop [0]{}%
\providecommand \bibitemNoStop [0]{.\EOS\space}%
\providecommand \EOS [0]{\spacefactor3000\relax}%
\providecommand \BibitemShut  [1]{\csname bibitem#1\endcsname}%
\let\auto@bib@innerbib\@empty
%</preamble>
\bibitem [{\citenamefont {Finegold}\ and\ \citenamefont
  {Donnell}(1979)}]{Finegold1979}%
  \BibitemOpen
  \bibfield  {author} {\bibinfo {author} {\bibfnamefont {L.}~\bibnamefont
  {Finegold}}\ and\ \bibinfo {author} {\bibfnamefont {J.~T.}\ \bibnamefont
  {Donnell}},\ }\bibfield  {title} {\bibinfo {title} {Maximum density of random
  placing of membrane particles},\ }\href {https://doi.org/10.1038/278443a0}
  {\bibfield  {journal} {\bibinfo  {journal} {Nature}\ }\textbf {\bibinfo
  {volume} {278}},\ \bibinfo {pages} {443} (\bibinfo {year}
  {1979})}\BibitemShut {NoStop}%
\bibitem [{\citenamefont {Balazs}\ and\ \citenamefont
  {Epstein}(1984)}]{Balaz1984}%
  \BibitemOpen
  \bibfield  {author} {\bibinfo {author} {\bibfnamefont {A.~C.}\ \bibnamefont
  {Balazs}}\ and\ \bibinfo {author} {\bibfnamefont {I.~R.}\ \bibnamefont
  {Epstein}},\ }\bibfield  {title} {\bibinfo {title} {Kinetics of irreversible
  dissociation for proteins bound cooperatively to {DNA}},\ }\href
  {https://doi.org/https://doi.org/10.1002/bip.360230709} {\bibfield  {journal}
  {\bibinfo  {journal} {Biopolymers}\ }\textbf {\bibinfo {volume} {23}},\
  \bibinfo {pages} {1249} (\bibinfo {year} {1984})}\BibitemShut {NoStop}%
\bibitem [{\citenamefont {Schwager}\ \emph {et~al.}(2008)\citenamefont
  {Schwager}, \citenamefont {Wolf},\ and\ \citenamefont
  {P\"oschel}}]{Schwager2008}%
  \BibitemOpen
  \bibfield  {author} {\bibinfo {author} {\bibfnamefont {T.}~\bibnamefont
  {Schwager}}, \bibinfo {author} {\bibfnamefont {D.~E.}\ \bibnamefont {Wolf}},\
  and\ \bibinfo {author} {\bibfnamefont {T.}~\bibnamefont {P\"oschel}},\
  }\bibfield  {title} {\bibinfo {title} {Fractal substructure of a
  nanopowder},\ }\href {https://doi.org/10.1103/PhysRevLett.100.218002}
  {\bibfield  {journal} {\bibinfo  {journal} {Phys. Rev. Lett.}\ }\textbf
  {\bibinfo {volume} {100}},\ \bibinfo {pages} {218002} (\bibinfo {year}
  {2008})}\BibitemShut {NoStop}%
\bibitem [{\citenamefont {Jensen}(1999)}]{Pablo1999}%
  \BibitemOpen
  \bibfield  {author} {\bibinfo {author} {\bibfnamefont {P.}~\bibnamefont
  {Jensen}},\ }\bibfield  {title} {\bibinfo {title} {Growth of nanostructures
  by cluster deposition: Experiments and simple models},\ }\href
  {https://doi.org/10.1103/RevModPhys.71.1695} {\bibfield  {journal} {\bibinfo
  {journal} {Rev. Mod. Phys.}\ }\textbf {\bibinfo {volume} {71}},\ \bibinfo
  {pages} {1695} (\bibinfo {year} {1999})}\BibitemShut {NoStop}%
\bibitem [{\citenamefont {Plawsky}\ \emph {et~al.}(2009)\citenamefont
  {Plawsky}, \citenamefont {Kim},\ and\ \citenamefont
  {Schubert}}]{Plawsky2009}%
  \BibitemOpen
  \bibfield  {author} {\bibinfo {author} {\bibfnamefont {J.~L.}\ \bibnamefont
  {Plawsky}}, \bibinfo {author} {\bibfnamefont {J.~K.}\ \bibnamefont {Kim}},\
  and\ \bibinfo {author} {\bibfnamefont {E.~F.}\ \bibnamefont {Schubert}},\
  }\bibfield  {title} {\bibinfo {title} {Engineered nanoporous and
  nanostructured films},\ }\href
  {https://doi.org/https://doi.org/10.1016/S1369-7021(09)70179-8} {\bibfield
  {journal} {\bibinfo  {journal} {Materials Today}\ }\textbf {\bibinfo {volume}
  {12}},\ \bibinfo {pages} {36} (\bibinfo {year} {2009})}\BibitemShut {NoStop}%
\bibitem [{\citenamefont {Joshi}\ \emph {et~al.}(2016)\citenamefont {Joshi},
  \citenamefont {Bargteil}, \citenamefont {Caciagli}, \citenamefont
  {Burelbach}, \citenamefont {Xing}, \citenamefont {Nunes}, \citenamefont
  {Pinto}, \citenamefont {Araújo}, \citenamefont {Brujic},\ and\ \citenamefont
  {Eiser}}]{Joshi2016}%
  \BibitemOpen
  \bibfield  {author} {\bibinfo {author} {\bibfnamefont {D.}~\bibnamefont
  {Joshi}}, \bibinfo {author} {\bibfnamefont {D.}~\bibnamefont {Bargteil}},
  \bibinfo {author} {\bibfnamefont {A.}~\bibnamefont {Caciagli}}, \bibinfo
  {author} {\bibfnamefont {J.}~\bibnamefont {Burelbach}}, \bibinfo {author}
  {\bibfnamefont {Z.}~\bibnamefont {Xing}}, \bibinfo {author} {\bibfnamefont
  {A.~S.}\ \bibnamefont {Nunes}}, \bibinfo {author} {\bibfnamefont {D.~E.~P.}\
  \bibnamefont {Pinto}}, \bibinfo {author} {\bibfnamefont {N.~A.~M.}\
  \bibnamefont {Araújo}}, \bibinfo {author} {\bibfnamefont {J.}~\bibnamefont
  {Brujic}},\ and\ \bibinfo {author} {\bibfnamefont {E.}~\bibnamefont
  {Eiser}},\ }\bibfield  {title} {\bibinfo {title} {Kinetic control of the
  coverage of oil droplets by dna-functionalized colloids},\ }\href
  {https://doi.org/10.1126/sciadv.1600881} {\bibfield  {journal} {\bibinfo
  {journal} {Science Advances}\ }\textbf {\bibinfo {volume} {2}},\ \bibinfo
  {pages} {e1600881} (\bibinfo {year} {2016})}\BibitemShut {NoStop}%
\bibitem [{\citenamefont {Evans}\ \emph {et~al.}(1983)\citenamefont {Evans},
  \citenamefont {Burgess},\ and\ \citenamefont {Hoffman}}]{Evans1983}%
  \BibitemOpen
  \bibfield  {author} {\bibinfo {author} {\bibfnamefont {J.~W.}\ \bibnamefont
  {Evans}}, \bibinfo {author} {\bibfnamefont {D.~R.}\ \bibnamefont {Burgess}},\
  and\ \bibinfo {author} {\bibfnamefont {D.~K.}\ \bibnamefont {Hoffman}},\
  }\bibfield  {title} {\bibinfo {title} {Irreversible random and cooperative
  processes on lattices: Exact and approximate hierarchy truncation and
  solution},\ }\href {https://doi.org/10.1063/1.445595} {\bibfield  {journal}
  {\bibinfo  {journal} {The Journal of Chemical Physics}\ }\textbf {\bibinfo
  {volume} {79}},\ \bibinfo {pages} {5011} (\bibinfo {year}
  {1983})}\BibitemShut {NoStop}%
\bibitem [{\citenamefont {Evans}(1993)}]{Evans1993}%
  \BibitemOpen
  \bibfield  {author} {\bibinfo {author} {\bibfnamefont {J.~W.}\ \bibnamefont
  {Evans}},\ }\bibfield  {title} {\bibinfo {title} {Random and cooperative
  sequential adsorption},\ }\href {https://doi.org/10.1103/RevModPhys.65.1281}
  {\bibfield  {journal} {\bibinfo  {journal} {Rev. Mod. Phys.}\ }\textbf
  {\bibinfo {volume} {65}},\ \bibinfo {pages} {1281} (\bibinfo {year}
  {1993})}\BibitemShut {NoStop}%
\bibitem [{\citenamefont {Schaaf}\ \emph {et~al.}(2000)\citenamefont {Schaaf},
  \citenamefont {Voegel},\ and\ \citenamefont {Senger}}]{Schaaf2000}%
  \BibitemOpen
  \bibfield  {author} {\bibinfo {author} {\bibfnamefont {P.}~\bibnamefont
  {Schaaf}}, \bibinfo {author} {\bibfnamefont {J.-C.}\ \bibnamefont {Voegel}},\
  and\ \bibinfo {author} {\bibfnamefont {B.}~\bibnamefont {Senger}},\
  }\bibfield  {title} {\bibinfo {title} {From random sequential adsorption to
  ballistic deposition: A general view of irreversible deposition processes},\
  }\href {https://doi.org/10.1021/jp9933065} {\bibfield  {journal} {\bibinfo
  {journal} {The Journal of Physical Chemistry B}\ }\textbf {\bibinfo {volume}
  {104}},\ \bibinfo {pages} {2204} (\bibinfo {year} {2000})}\BibitemShut
  {NoStop}%
\bibitem [{\citenamefont {Talbot}\ \emph {et~al.}(2000)\citenamefont {Talbot},
  \citenamefont {Tarjus}, \citenamefont {{Van Tassel}},\ and\ \citenamefont
  {Viot}}]{Talbrot2000}%
  \BibitemOpen
  \bibfield  {author} {\bibinfo {author} {\bibfnamefont {J.}~\bibnamefont
  {Talbot}}, \bibinfo {author} {\bibfnamefont {G.}~\bibnamefont {Tarjus}},
  \bibinfo {author} {\bibfnamefont {P.}~\bibnamefont {{Van Tassel}}},\ and\
  \bibinfo {author} {\bibfnamefont {P.}~\bibnamefont {Viot}},\ }\bibfield
  {title} {\bibinfo {title} {From car parking to protein adsorption: an
  overview of sequential adsorption processes},\ }\href
  {https://doi.org/https://doi.org/10.1016/S0927-7757(99)00409-4} {\bibfield
  {journal} {\bibinfo  {journal} {Colloids and Surfaces A: Physicochemical and
  Engineering Aspects}\ }\textbf {\bibinfo {volume} {165}},\ \bibinfo {pages}
  {287} (\bibinfo {year} {2000})}\BibitemShut {NoStop}%
\bibitem [{\citenamefont {Kubala}\ \emph {et~al.}(2022)\citenamefont {Kubala},
  \citenamefont {Batys}, \citenamefont {Barbasz}, \citenamefont {Weroński},\
  and\ \citenamefont {Cieśla}}]{Kubala2022}%
  \BibitemOpen
  \bibfield  {author} {\bibinfo {author} {\bibfnamefont {P.}~\bibnamefont
  {Kubala}}, \bibinfo {author} {\bibfnamefont {P.}~\bibnamefont {Batys}},
  \bibinfo {author} {\bibfnamefont {J.}~\bibnamefont {Barbasz}}, \bibinfo
  {author} {\bibfnamefont {P.}~\bibnamefont {Weroński}},\ and\ \bibinfo
  {author} {\bibfnamefont {M.}~\bibnamefont {Cieśla}},\ }\bibfield  {title}
  {\bibinfo {title} {Random sequential adsorption: An efficient tool for
  investigating the deposition of macromolecules and colloidal particles},\
  }\href {https://doi.org/https://doi.org/10.1016/j.cis.2022.102692} {\bibfield
   {journal} {\bibinfo  {journal} {Advances in Colloid and Interface Science}\
  }\textbf {\bibinfo {volume} {306}},\ \bibinfo {pages} {102692} (\bibinfo
  {year} {2022})}\BibitemShut {NoStop}%
\bibitem [{\citenamefont {Feng}\ \emph {et~al.}(2025)\citenamefont {Feng},
  \citenamefont {Del~Duca}, \citenamefont {Buffo},\ and\ \citenamefont
  {Simone}}]{Feng2025}%
  \BibitemOpen
  \bibfield  {author} {\bibinfo {author} {\bibfnamefont {Y.}~\bibnamefont
  {Feng}}, \bibinfo {author} {\bibfnamefont {G.}~\bibnamefont {Del~Duca}},
  \bibinfo {author} {\bibfnamefont {A.}~\bibnamefont {Buffo}},\ and\ \bibinfo
  {author} {\bibfnamefont {E.}~\bibnamefont {Simone}},\ }\bibfield  {title}
  {\bibinfo {title} {Random sequential adsorption of pickering particles onto
  spherical emulsion droplet surfaces},\ }\href
  {https://doi.org/10.1103/hkc7-5dfm} {\bibfield  {journal} {\bibinfo
  {journal} {Phys. Rev. E}\ }\textbf {\bibinfo {volume} {112}},\ \bibinfo
  {pages} {065506} (\bibinfo {year} {2025})}\BibitemShut {NoStop}%
\bibitem [{\citenamefont {Manna}\ and\ \citenamefont
  {Svrakic}(1991)}]{Manna1991}%
  \BibitemOpen
  \bibfield  {author} {\bibinfo {author} {\bibfnamefont {S.~S.}\ \bibnamefont
  {Manna}}\ and\ \bibinfo {author} {\bibfnamefont {N.~M.}\ \bibnamefont
  {Svrakic}},\ }\bibfield  {title} {\bibinfo {title} {Random sequential
  adsorption: line segments on the square lattice},\ }\href
  {https://doi.org/10.1088/0305-4470/24/12/003} {\bibfield  {journal} {\bibinfo
   {journal} {Journal of Physics A: Mathematical and General}\ }\textbf
  {\bibinfo {volume} {24}},\ \bibinfo {pages} {L671} (\bibinfo {year}
  {1991})}\BibitemShut {NoStop}%
\bibitem [{\citenamefont {Bonnier}\ \emph {et~al.}(1994)\citenamefont
  {Bonnier}, \citenamefont {Hontebeyrie}, \citenamefont {Leroyer},
  \citenamefont {Meyers},\ and\ \citenamefont {Pommiers}}]{Bonnnier1994}%
  \BibitemOpen
  \bibfield  {author} {\bibinfo {author} {\bibfnamefont {B.}~\bibnamefont
  {Bonnier}}, \bibinfo {author} {\bibfnamefont {M.}~\bibnamefont
  {Hontebeyrie}}, \bibinfo {author} {\bibfnamefont {Y.}~\bibnamefont
  {Leroyer}}, \bibinfo {author} {\bibfnamefont {C.}~\bibnamefont {Meyers}},\
  and\ \bibinfo {author} {\bibfnamefont {E.}~\bibnamefont {Pommiers}},\
  }\bibfield  {title} {\bibinfo {title} {Adsorption of line segments on a
  square lattice},\ }\href {https://doi.org/10.1103/PhysRevE.49.305} {\bibfield
   {journal} {\bibinfo  {journal} {Phys. Rev. E}\ }\textbf {\bibinfo {volume}
  {49}},\ \bibinfo {pages} {305} (\bibinfo {year} {1994})}\BibitemShut
  {NoStop}%
\bibitem [{\citenamefont {Kondrat}\ and\ \citenamefont
  {P\ifmmode~\mbox{\c{e}}\else \c{e}\fi{}kalski}(2001)}]{Kondrat2001}%
  \BibitemOpen
  \bibfield  {author} {\bibinfo {author} {\bibfnamefont {G.}~\bibnamefont
  {Kondrat}}\ and\ \bibinfo {author} {\bibfnamefont {A.}~\bibnamefont
  {P\ifmmode~\mbox{\c{e}}\else \c{e}\fi{}kalski}},\ }\bibfield  {title}
  {\bibinfo {title} {Percolation and jamming in random sequential adsorption of
  linear segments on a square lattice},\ }\href
  {https://doi.org/10.1103/PhysRevE.63.051108} {\bibfield  {journal} {\bibinfo
  {journal} {Phys. Rev. E}\ }\textbf {\bibinfo {volume} {63}},\ \bibinfo
  {pages} {051108} (\bibinfo {year} {2001})}\BibitemShut {NoStop}%
\bibitem [{\citenamefont {Lebovka}\ \emph {et~al.}(2011)\citenamefont
  {Lebovka}, \citenamefont {Karmazina}, \citenamefont {Tarasevich},\ and\
  \citenamefont {Laptev}}]{Lebovka2011}%
  \BibitemOpen
  \bibfield  {author} {\bibinfo {author} {\bibfnamefont {N.~I.}\ \bibnamefont
  {Lebovka}}, \bibinfo {author} {\bibfnamefont {N.~N.}\ \bibnamefont
  {Karmazina}}, \bibinfo {author} {\bibfnamefont {Y.~Y.}\ \bibnamefont
  {Tarasevich}},\ and\ \bibinfo {author} {\bibfnamefont {V.~V.}\ \bibnamefont
  {Laptev}},\ }\bibfield  {title} {\bibinfo {title} {Random sequential
  adsorption of partially oriented linear $k$-mers on a square lattice},\
  }\href {https://doi.org/10.1103/PhysRevE.84.061603} {\bibfield  {journal}
  {\bibinfo  {journal} {Phys. Rev. E}\ }\textbf {\bibinfo {volume} {84}},\
  \bibinfo {pages} {061603} (\bibinfo {year} {2011})}\BibitemShut {NoStop}%
\bibitem [{\citenamefont {Koza}\ and\ \citenamefont
  {Kondrat}(2025)}]{Koza2025}%
  \BibitemOpen
  \bibfield  {author} {\bibinfo {author} {\bibfnamefont {Z.}~\bibnamefont
  {Koza}}\ and\ \bibinfo {author} {\bibfnamefont {G.}~\bibnamefont {Kondrat}},\
  }\bibfield  {title} {\bibinfo {title} {Percolation and jamming in random
  sequential adsorption of straight $k$-mers on square, triangular, and cubic
  lattices},\ }\href {https://doi.org/10.1103/PhysRevE.111.034112} {\bibfield
  {journal} {\bibinfo  {journal} {Phys. Rev. E}\ }\textbf {\bibinfo {volume}
  {111}},\ \bibinfo {pages} {034112} (\bibinfo {year} {2025})}\BibitemShut
  {NoStop}%
\bibitem [{\citenamefont {Nakamura}(1986)}]{Nakamura1986}%
  \BibitemOpen
  \bibfield  {author} {\bibinfo {author} {\bibfnamefont {M.}~\bibnamefont
  {Nakamura}},\ }\bibfield  {title} {\bibinfo {title} {Random sequential
  packing in square cellular structures},\ }\href
  {https://doi.org/10.1088/0305-4470/19/12/020} {\bibfield  {journal} {\bibinfo
   {journal} {Journal of Physics A: Mathematical and General}\ }\textbf
  {\bibinfo {volume} {19}},\ \bibinfo {pages} {2345} (\bibinfo {year}
  {1986})}\BibitemShut {NoStop}%
\bibitem [{\citenamefont {Brosilow}\ \emph {et~al.}(1991)\citenamefont
  {Brosilow}, \citenamefont {Ziff},\ and\ \citenamefont
  {Vigil}}]{Borosilov1991}%
  \BibitemOpen
  \bibfield  {author} {\bibinfo {author} {\bibfnamefont {B.~J.}\ \bibnamefont
  {Brosilow}}, \bibinfo {author} {\bibfnamefont {R.~M.}\ \bibnamefont {Ziff}},\
  and\ \bibinfo {author} {\bibfnamefont {R.~D.}\ \bibnamefont {Vigil}},\
  }\bibfield  {title} {\bibinfo {title} {Random sequential adsorption of
  parallel squares},\ }\href {https://doi.org/10.1103/PhysRevA.43.631}
  {\bibfield  {journal} {\bibinfo  {journal} {Phys. Rev. A}\ }\textbf {\bibinfo
  {volume} {43}},\ \bibinfo {pages} {631} (\bibinfo {year} {1991})}\BibitemShut
  {NoStop}%
\bibitem [{\citenamefont {Ramirez-Pastor}\ \emph {et~al.}(2019)\citenamefont
  {Ramirez-Pastor}, \citenamefont {Centres}, \citenamefont {Vogel},\ and\
  \citenamefont {Vald\'es}}]{Ramirez-Square}%
  \BibitemOpen
  \bibfield  {author} {\bibinfo {author} {\bibfnamefont {A.~J.}\ \bibnamefont
  {Ramirez-Pastor}}, \bibinfo {author} {\bibfnamefont {P.~M.}\ \bibnamefont
  {Centres}}, \bibinfo {author} {\bibfnamefont {E.~E.}\ \bibnamefont {Vogel}},\
  and\ \bibinfo {author} {\bibfnamefont {J.~F.}\ \bibnamefont {Vald\'es}},\
  }\bibfield  {title} {\bibinfo {title} {Jamming and percolation for deposition
  of ${k}^{2}$-mers on square lattices: A monte carlo simulation study},\
  }\href {https://doi.org/10.1103/PhysRevE.99.042131} {\bibfield  {journal}
  {\bibinfo  {journal} {Phys. Rev. E}\ }\textbf {\bibinfo {volume} {99}},\
  \bibinfo {pages} {042131} (\bibinfo {year} {2019})}\BibitemShut {NoStop}%
\bibitem [{\citenamefont {Vigil}\ and\ \citenamefont {Ziff}(1989)}]{Ziff1989}%
  \BibitemOpen
  \bibfield  {author} {\bibinfo {author} {\bibfnamefont {R.~D.}\ \bibnamefont
  {Vigil}}\ and\ \bibinfo {author} {\bibfnamefont {R.~M.}\ \bibnamefont
  {Ziff}},\ }\bibfield  {title} {\bibinfo {title} {Random sequential adsorption
  of unoriented rectangles onto a plane},\ }\href
  {https://doi.org/10.1063/1.457021} {\bibfield  {journal} {\bibinfo  {journal}
  {The Journal of Chemical Physics}\ }\textbf {\bibinfo {volume} {91}},\
  \bibinfo {pages} {2599} (\bibinfo {year} {1989})}\BibitemShut {NoStop}%
\bibitem [{\citenamefont {Lebovka}\ \emph {et~al.}(2020)\citenamefont
  {Lebovka}, \citenamefont {Vygornitskii},\ and\ \citenamefont
  {Tarasevich}}]{Lebovka2020}%
  \BibitemOpen
  \bibfield  {author} {\bibinfo {author} {\bibfnamefont {N.~I.}\ \bibnamefont
  {Lebovka}}, \bibinfo {author} {\bibfnamefont {N.~V.}\ \bibnamefont
  {Vygornitskii}},\ and\ \bibinfo {author} {\bibfnamefont {Y.~Y.}\ \bibnamefont
  {Tarasevich}},\ }\bibfield  {title} {\bibinfo {title} {Random sequential
  adsorption of partially ordered discorectangles onto a continuous plane},\
  }\href {https://doi.org/10.1103/PhysRevE.102.022133} {\bibfield  {journal}
  {\bibinfo  {journal} {Phys. Rev. E}\ }\textbf {\bibinfo {volume} {102}},\
  \bibinfo {pages} {022133} (\bibinfo {year} {2020})}\BibitemShut {NoStop}%
\bibitem [{\citenamefont {Petrone}\ and\ \citenamefont
  {Cie\ifmmode~\acute{s}\else \'{s}\fi{}la}(2021)}]{Petrone2021}%
  \BibitemOpen
  \bibfield  {author} {\bibinfo {author} {\bibfnamefont {L.}~\bibnamefont
  {Petrone}}\ and\ \bibinfo {author} {\bibfnamefont {M.}~\bibnamefont
  {Cie\ifmmode~\acute{s}\else \'{s}\fi{}la}},\ }\bibfield  {title} {\bibinfo
  {title} {Random sequential adsorption of oriented rectangles with random
  aspect ratio},\ }\href {https://doi.org/10.1103/PhysRevE.104.034903}
  {\bibfield  {journal} {\bibinfo  {journal} {Phys. Rev. E}\ }\textbf {\bibinfo
  {volume} {104}},\ \bibinfo {pages} {034903} (\bibinfo {year}
  {2021})}\BibitemShut {NoStop}%
\bibitem [{\citenamefont {Feder}(1980)}]{Feder1980}%
  \BibitemOpen
  \bibfield  {author} {\bibinfo {author} {\bibfnamefont {J.}~\bibnamefont
  {Feder}},\ }\bibfield  {title} {\bibinfo {title} {Random sequential
  adsorption},\ }\href
  {https://doi.org/https://doi.org/10.1016/0022-5193(80)90358-6} {\bibfield
  {journal} {\bibinfo  {journal} {Journal of Theoretical Biology}\ }\textbf
  {\bibinfo {volume} {87}},\ \bibinfo {pages} {237} (\bibinfo {year}
  {1980})}\BibitemShut {NoStop}%
\bibitem [{\citenamefont {Onoda}\ and\ \citenamefont
  {Liniger}(1986)}]{Onoda1986}%
  \BibitemOpen
  \bibfield  {author} {\bibinfo {author} {\bibfnamefont {G.~Y.}\ \bibnamefont
  {Onoda}}\ and\ \bibinfo {author} {\bibfnamefont {E.~G.}\ \bibnamefont
  {Liniger}},\ }\bibfield  {title} {\bibinfo {title} {Experimental
  determination of the random-parking limit in two dimensions},\ }\href
  {https://doi.org/10.1103/PhysRevA.33.715} {\bibfield  {journal} {\bibinfo
  {journal} {Phys. Rev. A}\ }\textbf {\bibinfo {volume} {33}},\ \bibinfo
  {pages} {715} (\bibinfo {year} {1986})}\BibitemShut {NoStop}%
\bibitem [{\citenamefont {Sherwood}(1990)}]{Sherwood1990}%
  \BibitemOpen
  \bibfield  {author} {\bibinfo {author} {\bibfnamefont {J.~D.}\ \bibnamefont
  {Sherwood}},\ }\bibfield  {title} {\bibinfo {title} {Random sequential
  adsorption of lines and ellipses},\ }\href
  {https://doi.org/10.1088/0305-4470/23/13/021} {\bibfield  {journal} {\bibinfo
   {journal} {Journal of Physics A: Mathematical and General}\ }\textbf
  {\bibinfo {volume} {23}},\ \bibinfo {pages} {2827} (\bibinfo {year}
  {1990})}\BibitemShut {NoStop}%
\bibitem [{\citenamefont {Viot}\ \emph {et~al.}(1992)\citenamefont {Viot},
  \citenamefont {Tarjus}, \citenamefont {Ricci},\ and\ \citenamefont
  {Talbot}}]{Viot1992}%
  \BibitemOpen
  \bibfield  {author} {\bibinfo {author} {\bibfnamefont {P.}~\bibnamefont
  {Viot}}, \bibinfo {author} {\bibfnamefont {G.}~\bibnamefont {Tarjus}},
  \bibinfo {author} {\bibfnamefont {S.~M.}\ \bibnamefont {Ricci}},\ and\
  \bibinfo {author} {\bibfnamefont {J.}~\bibnamefont {Talbot}},\ }\bibfield
  {title} {\bibinfo {title} {Random sequential adsorption of anisotropic
  particles. i. jamming limit and asymptotic behavior},\ }\href
  {https://doi.org/10.1063/1.463820} {\bibfield  {journal} {\bibinfo  {journal}
  {The Journal of Chemical Physics}\ }\textbf {\bibinfo {volume} {97}},\
  \bibinfo {pages} {5212} (\bibinfo {year} {1992})}\BibitemShut {NoStop}%
\bibitem [{\citenamefont {Cieśla}\ \emph {et~al.}(2016)\citenamefont
  {Cieśla}, \citenamefont {Paja̧k},\ and\ \citenamefont {Ziff}}]{Ciesla2016}%
  \BibitemOpen
  \bibfield  {author} {\bibinfo {author} {\bibfnamefont {M.}~\bibnamefont
  {Cie\'sla}}, \bibinfo {author} {\bibfnamefont {G.}~\bibnamefont {Paj\c{a}k}},\
  and\ \bibinfo {author} {\bibfnamefont {R.~M.}\ \bibnamefont {Ziff}},\
  }\bibfield  {title} {\bibinfo {title} {In a search for a shape maximizing
  packing fraction for two-dimensional random sequential adsorption},\ }\href
  {https://doi.org/10.1063/1.4959584} {\bibfield  {journal} {\bibinfo
  {journal} {The Journal of Chemical Physics}\ }\textbf {\bibinfo {volume}
  {145}},\ \bibinfo {pages} {044708} (\bibinfo {year} {2016})}\BibitemShut
  {NoStop}%
\bibitem [{\citenamefont {Meakin}\ and\ \citenamefont
  {Jullien}(1992)}]{Meakin1992}%
  \BibitemOpen
  \bibfield  {author} {\bibinfo {author} {\bibfnamefont {P.}~\bibnamefont
  {Meakin}}\ and\ \bibinfo {author} {\bibfnamefont {R.}~\bibnamefont
  {Jullien}},\ }\bibfield  {title} {\bibinfo {title} {Random-sequential
  adsorption of disks of different sizes},\ }\href
  {https://doi.org/10.1103/PhysRevA.46.2029} {\bibfield  {journal} {\bibinfo
  {journal} {Phys. Rev. A}\ }\textbf {\bibinfo {volume} {46}},\ \bibinfo
  {pages} {2029} (\bibinfo {year} {1992})}\BibitemShut {NoStop}%
\bibitem [{\citenamefont {Brilliantov}\ \emph {et~al.}(1996)\citenamefont
  {Brilliantov}, \citenamefont {Andrienko}, \citenamefont {Krapivsky},\ and\
  \citenamefont {Kurths}}]{Brilliantov1996}%
  \BibitemOpen
  \bibfield  {author} {\bibinfo {author} {\bibfnamefont {N.~V.}\ \bibnamefont
  {Brilliantov}}, \bibinfo {author} {\bibfnamefont {Y.~A.}\ \bibnamefont
  {Andrienko}}, \bibinfo {author} {\bibfnamefont {P.~L.}\ \bibnamefont
  {Krapivsky}},\ and\ \bibinfo {author} {\bibfnamefont {J.}~\bibnamefont
  {Kurths}},\ }\bibfield  {title} {\bibinfo {title} {Fractal formation and
  ordering in random sequential adsorption},\ }\href
  {https://doi.org/10.1103/PhysRevLett.76.4058} {\bibfield  {journal} {\bibinfo
   {journal} {Phys. Rev. Lett.}\ }\textbf {\bibinfo {volume} {76}},\ \bibinfo
  {pages} {4058} (\bibinfo {year} {1996})}\BibitemShut {NoStop}%
\bibitem [{\citenamefont {Hart}\ and\ \citenamefont {Aar\~ao
  Reis}(2016)}]{Hart2016}%
  \BibitemOpen
  \bibfield  {author} {\bibinfo {author} {\bibfnamefont {R.~C.}\ \bibnamefont
  {Hart}}\ and\ \bibinfo {author} {\bibfnamefont {F.~D.~A.}\ \bibnamefont
  {Aar\~ao Reis}},\ }\bibfield  {title} {\bibinfo {title} {Random sequential
  adsorption of polydisperse mixtures on lattices},\ }\href
  {https://doi.org/10.1103/PhysRevE.94.022802} {\bibfield  {journal} {\bibinfo
  {journal} {Phys. Rev. E}\ }\textbf {\bibinfo {volume} {94}},\ \bibinfo
  {pages} {022802} (\bibinfo {year} {2016})}\BibitemShut {NoStop}%
\bibitem [{\citenamefont {Wagaskar}\ \emph {et~al.}(2020)\citenamefont
  {Wagaskar}, \citenamefont {Late}, \citenamefont {Banpurkar}, \citenamefont
  {Limaye},\ and\ \citenamefont {Shelke}}]{Wagaskar2020}%
  \BibitemOpen
  \bibfield  {author} {\bibinfo {author} {\bibfnamefont {K.~V.}\ \bibnamefont
  {Wagaskar}}, \bibinfo {author} {\bibfnamefont {R.}~\bibnamefont {Late}},
  \bibinfo {author} {\bibfnamefont {A.~G.}\ \bibnamefont {Banpurkar}}, \bibinfo
  {author} {\bibfnamefont {A.~V.}\ \bibnamefont {Limaye}},\ and\ \bibinfo
  {author} {\bibfnamefont {P.~B.}\ \bibnamefont {Shelke}},\ }\bibfield  {title}
  {\bibinfo {title} {Simulation studies of random sequential adsorption (rsa)
  of mixture of two-component circular discs},\ }\href
  {https://doi.org/10.1007/s10955-020-02660-7} {\bibfield  {journal} {\bibinfo
  {journal} {Journal of Statistical Physics}\ }\textbf {\bibinfo {volume}
  {181}},\ \bibinfo {pages} {2191} (\bibinfo {year} {2020})}\BibitemShut
  {NoStop}%
\bibitem [{\citenamefont {Kundu}\ \emph {et~al.}(2022)\citenamefont {Kundu},
  \citenamefont {Prates},\ and\ \citenamefont {Araújo}}]{Kundu2022}%
  \BibitemOpen
  \bibfield  {author} {\bibinfo {author} {\bibfnamefont {S.}~\bibnamefont
  {Kundu}}, \bibinfo {author} {\bibfnamefont {H.~C.}\ \bibnamefont {Prates}},\
  and\ \bibinfo {author} {\bibfnamefont {N.~A.~M.}\ \bibnamefont {Araújo}},\
  }\bibfield  {title} {\bibinfo {title} {Jamming and percolation in the random
  sequential adsorption of a binary mixture on the square lattice},\ }\href
  {https://doi.org/10.1088/1751-8121/ac6241} {\bibfield  {journal} {\bibinfo
  {journal} {Journal of Physics A: Mathematical and Theoretical}\ }\textbf
  {\bibinfo {volume} {55}},\ \bibinfo {pages} {204005} (\bibinfo {year}
  {2022})}\BibitemShut {NoStop}%
\bibitem [{\citenamefont {Mandelbrot}(1982)}]{Mandelbrot1982}%
  \BibitemOpen
  \bibfield  {author} {\bibinfo {author} {\bibfnamefont {B.~B.}\ \bibnamefont
  {Mandelbrot}},\ }\href@noop {} {\emph {\bibinfo {title} {The Fractal Geometry
  of Nature}}}\ (\bibinfo  {publisher} {W. H. Freeman and Company},\ \bibinfo
  {address} {San Francisco, USA},\ \bibinfo {year} {1982})\BibitemShut
  {NoStop}%
\bibitem [{\citenamefont {Feder}(1988)}]{Feder1988}%
  \BibitemOpen
  \bibfield  {author} {\bibinfo {author} {\bibfnamefont {J.}~\bibnamefont
  {Feder}},\ }\href@noop {} {\emph {\bibinfo {title} {Fractals}}}\ (\bibinfo
  {publisher} {Plenum Press},\ \bibinfo {address} {New York, USA},\ \bibinfo
  {year} {1988})\BibitemShut {NoStop}%
\bibitem [{\citenamefont {Ball}(2009)}]{Ball2009}%
  \BibitemOpen
  \bibfield  {author} {\bibinfo {author} {\bibfnamefont {P.}~\bibnamefont
  {Ball}},\ }\href@noop {} {\emph {\bibinfo {title} {Branches}}}\ (\bibinfo
  {publisher} {Oxford University Press},\ \bibinfo {address} {Oxford, UK},\
  \bibinfo {year} {2009})\BibitemShut {NoStop}%
\bibitem [{\citenamefont {Kozicki}(2021)}]{Kozicki2021}%
  \BibitemOpen
  \bibfield  {author} {\bibinfo {author} {\bibfnamefont {M.~N.}\ \bibnamefont
  {Kozicki}},\ }\bibfield  {title} {\bibinfo {title} {Information in
  electrodeposited dendrites},\ }\href
  {https://doi.org/10.1080/23746149.2021.1920846} {\bibfield  {journal}
  {\bibinfo  {journal} {Advances in Physics: X}\ }\textbf {\bibinfo {volume}
  {6}},\ \bibinfo {pages} {1920846} (\bibinfo {year} {2021})}\BibitemShut
  {NoStop}%
\bibitem [{\citenamefont {Cartwright}\ \emph {et~al.}(2026)\citenamefont
  {Cartwright}, \citenamefont {Cockell}, \citenamefont {Goehring},
  \citenamefont {Holler}, \citenamefont {Jordan}, \citenamefont {Knoll},
  \citenamefont {Kotopoulou}, \citenamefont {Loron}, \citenamefont {McMahon},
  \citenamefont {Morris}, \citenamefont {Neubeck}, \citenamefont {Pimentel},
  \citenamefont {Sainz-Díaz}, \citenamefont {Shahidzadeh},\ and\ \citenamefont
  {Szymczak}}]{Morris2026}%
  \BibitemOpen
  \bibfield  {author} {\bibinfo {author} {\bibfnamefont {J.~H.}\ \bibnamefont
  {Cartwright}}, \bibinfo {author} {\bibfnamefont {C.~S.}\ \bibnamefont
  {Cockell}}, \bibinfo {author} {\bibfnamefont {L.}~\bibnamefont {Goehring}},
  \bibinfo {author} {\bibfnamefont {S.}~\bibnamefont {Holler}}, \bibinfo
  {author} {\bibfnamefont {S.~F.}\ \bibnamefont {Jordan}}, \bibinfo {author}
  {\bibfnamefont {P.}~\bibnamefont {Knoll}}, \bibinfo {author} {\bibfnamefont
  {E.}~\bibnamefont {Kotopoulou}}, \bibinfo {author} {\bibfnamefont {C.~C.}\
  \bibnamefont {Loron}}, \bibinfo {author} {\bibfnamefont {S.}~\bibnamefont
  {McMahon}}, \bibinfo {author} {\bibfnamefont {S.~W.}\ \bibnamefont {Morris}},
  \bibinfo {author} {\bibfnamefont {A.}~\bibnamefont {Neubeck}}, \bibinfo
  {author} {\bibfnamefont {C.}~\bibnamefont {Pimentel}}, \bibinfo {author}
  {\bibfnamefont {C.~I.}\ \bibnamefont {Sainz-Díaz}}, \bibinfo {author}
  {\bibfnamefont {N.}~\bibnamefont {Shahidzadeh}},\ and\ \bibinfo {author}
  {\bibfnamefont {P.}~\bibnamefont {Szymczak}},\ }\bibfield  {title} {\bibinfo
  {title} {Self-organized pattern formation in geological soft matter},\ }\href
  {https://doi.org/https://doi.org/10.1016/j.physrep.2026.03.004} {\bibfield
  {journal} {\bibinfo  {journal} {Physics Reports}\ }\textbf {\bibinfo {volume}
  {1183}},\ \bibinfo {pages} {1} (\bibinfo {year} {2026})}\BibitemShut
  {NoStop}%
\bibitem [{\citenamefont {Chopard}\ \emph {et~al.}(1991)\citenamefont
  {Chopard}, \citenamefont {Herrmann},\ and\ \citenamefont
  {Vicsek}}]{Vicsek1991}%
  \BibitemOpen
  \bibfield  {author} {\bibinfo {author} {\bibfnamefont {B.}~\bibnamefont
  {Chopard}}, \bibinfo {author} {\bibfnamefont {H.~J.}\ \bibnamefont
  {Herrmann}},\ and\ \bibinfo {author} {\bibfnamefont {T.}~\bibnamefont
  {Vicsek}},\ }\bibfield  {title} {\bibinfo {title} {Structure and growth
  mechanism of mineral dendrites},\ }\href {https://doi.org/10.1038/353409a0}
  {\bibfield  {journal} {\bibinfo  {journal} {Nature}\ }\textbf {\bibinfo
  {volume} {353}},\ \bibinfo {pages} {409} (\bibinfo {year}
  {1991})}\BibitemShut {NoStop}%
\bibitem [{\citenamefont {Rodriguez-Iturbe}\ \emph {et~al.}(1992)\citenamefont
  {Rodriguez-Iturbe}, \citenamefont {Rinaldo}, \citenamefont {Rigon},
  \citenamefont {Bras}, \citenamefont {Ijjasz-Vasquez},\ and\ \citenamefont
  {Marani}}]{Ignacio1992}%
  \BibitemOpen
  \bibfield  {author} {\bibinfo {author} {\bibfnamefont {I.}~\bibnamefont
  {Rodriguez-Iturbe}}, \bibinfo {author} {\bibfnamefont {A.}~\bibnamefont
  {Rinaldo}}, \bibinfo {author} {\bibfnamefont {R.}~\bibnamefont {Rigon}},
  \bibinfo {author} {\bibfnamefont {R.~L.}\ \bibnamefont {Bras}}, \bibinfo
  {author} {\bibfnamefont {E.}~\bibnamefont {Ijjasz-Vasquez}},\ and\ \bibinfo
  {author} {\bibfnamefont {A.}~\bibnamefont {Marani}},\ }\bibfield  {title}
  {\bibinfo {title} {Fractal structures as least energy patterns: The case of
  river networks},\ }\href {https://doi.org/https://doi.org/10.1029/92GL00938}
  {\bibfield  {journal} {\bibinfo  {journal} {Geophysical Research Letters}\
  }\textbf {\bibinfo {volume} {19}},\ \bibinfo {pages} {889} (\bibinfo {year}
  {1992})}\BibitemShut {NoStop}%
\bibitem [{\citenamefont {Br{\'e}chignac}\ \emph {et~al.}(2003)\citenamefont
  {Br{\'e}chignac}, \citenamefont {Cahuzac}, \citenamefont {Carlier},
  \citenamefont {Colliex}, \citenamefont {de~Frutos}, \citenamefont
  {K{\'e}ba{\"\i}li}, \citenamefont {Le~Roux}, \citenamefont {Masson},\ and\
  \citenamefont {Yoon}}]{Cahuzac2003}%
  \BibitemOpen
  \bibfield  {author} {\bibinfo {author} {\bibfnamefont {C.}~\bibnamefont
  {Br{\'e}chignac}}, \bibinfo {author} {\bibfnamefont {P.}~\bibnamefont
  {Cahuzac}}, \bibinfo {author} {\bibfnamefont {F.}~\bibnamefont {Carlier}},
  \bibinfo {author} {\bibfnamefont {C.}~\bibnamefont {Colliex}}, \bibinfo
  {author} {\bibfnamefont {M.}~\bibnamefont {de~Frutos}}, \bibinfo {author}
  {\bibfnamefont {N.}~\bibnamefont {K{\'e}ba{\"\i}li}}, \bibinfo {author}
  {\bibfnamefont {J.}~\bibnamefont {Le~Roux}}, \bibinfo {author} {\bibfnamefont
  {A.}~\bibnamefont {Masson}},\ and\ \bibinfo {author} {\bibfnamefont
  {B.}~\bibnamefont {Yoon}},\ }\bibfield  {title} {\bibinfo {title} {Thermal
  and chemical nanofractalrelaxation},\ }\href
  {https://doi.org/10.1140/epjd/e2003-00159-8} {\bibfield  {journal} {\bibinfo
  {journal} {The European Physical Journal D - Atomic, Molecular, Optical and
  Plasma Physics}\ }\textbf {\bibinfo {volume} {24}},\ \bibinfo {pages} {265}
  (\bibinfo {year} {2003})}\BibitemShut {NoStop}%
\bibitem [{\citenamefont {Lando}\ \emph {et~al.}(2006)\citenamefont {Lando},
  \citenamefont {K\'eba\"{\i}li}, \citenamefont {Cahuzac}, \citenamefont
  {Masson},\ and\ \citenamefont {Br\'echignac}}]{Lando2006}%
  \BibitemOpen
  \bibfield  {author} {\bibinfo {author} {\bibfnamefont {A.}~\bibnamefont
  {Lando}}, \bibinfo {author} {\bibfnamefont {N.}~\bibnamefont
  {K\'eba\"{\i}li}}, \bibinfo {author} {\bibfnamefont {P.}~\bibnamefont
  {Cahuzac}}, \bibinfo {author} {\bibfnamefont {A.}~\bibnamefont {Masson}},\
  and\ \bibinfo {author} {\bibfnamefont {C.}~\bibnamefont {Br\'echignac}},\
  }\bibfield  {title} {\bibinfo {title} {Coarsening and pearling instabilities
  in silver nanofractal aggregates},\ }\href
  {https://doi.org/10.1103/PhysRevLett.97.133402} {\bibfield  {journal}
  {\bibinfo  {journal} {Phys. Rev. Lett.}\ }\textbf {\bibinfo {volume} {97}},\
  \bibinfo {pages} {133402} (\bibinfo {year} {2006})}\BibitemShut {NoStop}%
\bibitem [{\citenamefont {Dick}\ \emph {et~al.}(2011)\citenamefont {Dick},
  \citenamefont {Solov'yov},\ and\ \citenamefont {Solov'yov}}]{Veronika2011}%
  \BibitemOpen
  \bibfield  {author} {\bibinfo {author} {\bibfnamefont {V.~V.}\ \bibnamefont
  {Dick}}, \bibinfo {author} {\bibfnamefont {I.~A.}\ \bibnamefont
  {Solov'yov}},\ and\ \bibinfo {author} {\bibfnamefont {A.~V.}\ \bibnamefont
  {Solov'yov}},\ }\bibfield  {title} {\bibinfo {title} {Fragmentation pathways
  of nanofractal structures on surfaces},\ }\href
  {https://doi.org/10.1103/PhysRevB.84.115408} {\bibfield  {journal} {\bibinfo
  {journal} {Phys. Rev. B}\ }\textbf {\bibinfo {volume} {84}},\ \bibinfo
  {pages} {115408} (\bibinfo {year} {2011})}\BibitemShut {NoStop}%
\bibitem [{\citenamefont {Fairbanks}\ \emph {et~al.}(2011)\citenamefont
  {Fairbanks}, \citenamefont {McCarthy}, \citenamefont {Scott}, \citenamefont
  {Brown},\ and\ \citenamefont {Taylor}}]{Fairbanks2011}%
  \BibitemOpen
  \bibfield  {author} {\bibinfo {author} {\bibfnamefont {M.~S.}\ \bibnamefont
  {Fairbanks}}, \bibinfo {author} {\bibfnamefont {D.~N.}\ \bibnamefont
  {McCarthy}}, \bibinfo {author} {\bibfnamefont {S.~A.}\ \bibnamefont {Scott}},
  \bibinfo {author} {\bibfnamefont {S.~A.}\ \bibnamefont {Brown}},\ and\
  \bibinfo {author} {\bibfnamefont {R.~P.}\ \bibnamefont {Taylor}},\ }\bibfield
   {title} {\bibinfo {title} {Fractal electronic devices: simulation and
  implementation},\ }\href {https://doi.org/10.1088/0957-4484/22/36/365304}
  {\bibfield  {journal} {\bibinfo  {journal} {Nanotechnology}\ }\textbf
  {\bibinfo {volume} {22}},\ \bibinfo {pages} {365304} (\bibinfo {year}
  {2011})}\BibitemShut {NoStop}%
\bibitem [{\citenamefont {Becker}\ \emph {et~al.}(2019)\citenamefont {Becker},
  \citenamefont {Dantzig}, \citenamefont {Kolbe}, \citenamefont {Wiese},\ and\
  \citenamefont {Kargl}}]{Becker2019}%
  \BibitemOpen
  \bibfield  {author} {\bibinfo {author} {\bibfnamefont {M.}~\bibnamefont
  {Becker}}, \bibinfo {author} {\bibfnamefont {J.}~\bibnamefont {Dantzig}},
  \bibinfo {author} {\bibfnamefont {M.}~\bibnamefont {Kolbe}}, \bibinfo
  {author} {\bibfnamefont {S.}~\bibnamefont {Wiese}},\ and\ \bibinfo {author}
  {\bibfnamefont {F.}~\bibnamefont {Kargl}},\ }\bibfield  {title} {\bibinfo
  {title} {Dendrite orientation transition in alge alloys},\ }\href
  {https://doi.org/https://doi.org/10.1016/j.actamat.2018.12.001} {\bibfield
  {journal} {\bibinfo  {journal} {Acta Materialia}\ }\textbf {\bibinfo {volume}
  {165}},\ \bibinfo {pages} {666} (\bibinfo {year} {2019})}\BibitemShut
  {NoStop}%
\bibitem [{\citenamefont {Becker}\ \emph {et~al.}(2023)\citenamefont {Becker},
  \citenamefont {Wegener}, \citenamefont {Drescher},\ and\ \citenamefont
  {Kargl}}]{Becker2023}%
  \BibitemOpen
  \bibfield  {author} {\bibinfo {author} {\bibfnamefont {M.}~\bibnamefont
  {Becker}}, \bibinfo {author} {\bibfnamefont {M.}~\bibnamefont {Wegener}},
  \bibinfo {author} {\bibfnamefont {J.}~\bibnamefont {Drescher}},\ and\
  \bibinfo {author} {\bibfnamefont {F.}~\bibnamefont {Kargl}},\ }\bibfield
  {title} {\bibinfo {title} {Nucleation and growth dynamics of equiaxed
  dendrites in thin metallic al--cu and al--ge samples in microgravity and on
  earth},\ }\href {https://doi.org/10.1007/s11661-023-07079-9} {\bibfield
  {journal} {\bibinfo  {journal} {Metallurgical and Materials Transactions A}\
  }\textbf {\bibinfo {volume} {54}},\ \bibinfo {pages} {4188} (\bibinfo {year}
  {2023})}\BibitemShut {NoStop}%
\bibitem [{\citenamefont {Wang}\ and\ \citenamefont {Pandey}(1996)}]{Wang1996}%
  \BibitemOpen
  \bibfield  {author} {\bibinfo {author} {\bibfnamefont {J.-S.}\ \bibnamefont
  {Wang}}\ and\ \bibinfo {author} {\bibfnamefont {R.~B.}\ \bibnamefont
  {Pandey}},\ }\bibfield  {title} {\bibinfo {title} {Kinetics and jamming
  coverage in a random sequential adsorption of polymer chains},\ }\href
  {https://doi.org/10.1103/PhysRevLett.77.1773} {\bibfield  {journal} {\bibinfo
   {journal} {Phys. Rev. Lett.}\ }\textbf {\bibinfo {volume} {77}},\ \bibinfo
  {pages} {1773} (\bibinfo {year} {1996})}\BibitemShut {NoStop}%
\bibitem [{\citenamefont {Budinski-Petković}\ and\ \citenamefont
  {Kozmidis-Luburić}(1997)}]{Budinski1996}%
  \BibitemOpen
  \bibfield  {author} {\bibinfo {author} {\bibfnamefont {L.}~\bibnamefont
  {Budinski-Petković}}\ and\ \bibinfo {author} {\bibfnamefont
  {U.}~\bibnamefont {Kozmidis-Luburić}},\ }\bibfield  {title} {\bibinfo
  {title} {Jamming configurations for irreversible deposition on a square
  lattice},\ }\href
  {https://doi.org/https://doi.org/10.1016/S0378-4371(96)00374-3} {\bibfield
  {journal} {\bibinfo  {journal} {Physica A: Statistical Mechanics and its
  Applications}\ }\textbf {\bibinfo {volume} {236}},\ \bibinfo {pages} {211}
  (\bibinfo {year} {1997})}\BibitemShut {NoStop}%
\bibitem [{\citenamefont {Adamczyk}\ \emph {et~al.}(2008)\citenamefont
  {Adamczyk}, \citenamefont {Romiszowski},\ and\ \citenamefont
  {Sikorski}}]{Adamczyk2008}%
  \BibitemOpen
  \bibfield  {author} {\bibinfo {author} {\bibfnamefont {P.}~\bibnamefont
  {Adamczyk}}, \bibinfo {author} {\bibfnamefont {P.}~\bibnamefont
  {Romiszowski}},\ and\ \bibinfo {author} {\bibfnamefont {A.}~\bibnamefont
  {Sikorski}},\ }\bibfield  {title} {\bibinfo {title} {A simple model of stiff
  and flexible polymer chain adsorption: The influence of the internal chain
  architecture},\ }\href {https://doi.org/10.1063/1.2907715} {\bibfield
  {journal} {\bibinfo  {journal} {The Journal of Chemical Physics}\ }\textbf
  {\bibinfo {volume} {128}},\ \bibinfo {pages} {154911} (\bibinfo {year}
  {2008})}\BibitemShut {NoStop}%
\bibitem [{\citenamefont {Budinski-Petkovi\ifmmode~\acute{c}\else \'{c}\fi{}}\
  \emph {et~al.}(2008)\citenamefont {Budinski-Petkovi\ifmmode~\acute{c}\else
  \'{c}\fi{}}, \citenamefont {Vrhovac},\ and\ \citenamefont {Lon\ifmmode
  \check{c}\else \v{c}\fi{}arevi\ifmmode~\acute{c}\else
  \'{c}\fi{}}}]{Petkovic2008}%
  \BibitemOpen
  \bibfield  {author} {\bibinfo {author} {\bibfnamefont {L.}~\bibnamefont
  {Budinski-Petkovi\ifmmode~\acute{c}\else \'{c}\fi{}}}, \bibinfo {author}
  {\bibfnamefont {S.~B.}\ \bibnamefont {Vrhovac}},\ and\ \bibinfo {author}
  {\bibfnamefont {I.}~\bibnamefont {Lon\ifmmode \check{c}\else
  \v{c}\fi{}arevi\ifmmode~\acute{c}\else \'{c}\fi{}}},\ }\bibfield  {title}
  {\bibinfo {title} {Random sequential adsorption of polydisperse mixtures on
  discrete substrates},\ }\href {https://doi.org/10.1103/PhysRevE.78.061603}
  {\bibfield  {journal} {\bibinfo  {journal} {Phys. Rev. E}\ }\textbf {\bibinfo
  {volume} {78}},\ \bibinfo {pages} {061603} (\bibinfo {year}
  {2008})}\BibitemShut {NoStop}%
\bibitem [{\citenamefont {Ramirez}\ \emph {et~al.}(2023)\citenamefont
  {Ramirez}, \citenamefont {Pasinetti},\ and\ \citenamefont
  {Ramirez-Pastor}}]{Ramirez2023}%
  \BibitemOpen
  \bibfield  {author} {\bibinfo {author} {\bibfnamefont {L.~S.}\ \bibnamefont
  {Ramirez}}, \bibinfo {author} {\bibfnamefont {P.~M.}\ \bibnamefont
  {Pasinetti}},\ and\ \bibinfo {author} {\bibfnamefont {A.~J.}\ \bibnamefont
  {Ramirez-Pastor}},\ }\bibfield  {title} {\bibinfo {title} {Random sequential
  adsorption of self-avoiding chains on two-dimensional lattices},\ }\href
  {https://doi.org/10.1103/PhysRevE.107.064106} {\bibfield  {journal} {\bibinfo
   {journal} {Phys. Rev. E}\ }\textbf {\bibinfo {volume} {107}},\ \bibinfo
  {pages} {064106} (\bibinfo {year} {2023})}\BibitemShut {NoStop}%
\bibitem [{\citenamefont {Witten}\ and\ \citenamefont
  {Sander}(1981)}]{Witten1981}%
  \BibitemOpen
  \bibfield  {author} {\bibinfo {author} {\bibfnamefont {T.~A.}\ \bibnamefont
  {Witten}}\ and\ \bibinfo {author} {\bibfnamefont {L.~M.}\ \bibnamefont
  {Sander}},\ }\bibfield  {title} {\bibinfo {title} {Diffusion-limited
  aggregation, a kinetic critical phenomenon},\ }\href
  {https://doi.org/10.1103/PhysRevLett.47.1400} {\bibfield  {journal} {\bibinfo
   {journal} {Phys. Rev. Lett.}\ }\textbf {\bibinfo {volume} {47}},\ \bibinfo
  {pages} {1400} (\bibinfo {year} {1981})}\BibitemShut {NoStop}%
\bibitem [{\citenamefont {Muthukumar}(1983)}]{Muthukumar1983}%
  \BibitemOpen
  \bibfield  {author} {\bibinfo {author} {\bibfnamefont {M.}~\bibnamefont
  {Muthukumar}},\ }\bibfield  {title} {\bibinfo {title} {Mean-field theory for
  diffusion-limited cluster formation},\ }\href
  {https://doi.org/10.1103/PhysRevLett.50.839} {\bibfield  {journal} {\bibinfo
  {journal} {Phys. Rev. Lett.}\ }\textbf {\bibinfo {volume} {50}},\ \bibinfo
  {pages} {839} (\bibinfo {year} {1983})}\BibitemShut {NoStop}%
\bibitem [{\citenamefont {Tokuyama}\ and\ \citenamefont
  {Kawasaki}(1984)}]{Tokuyama1984}%
  \BibitemOpen
  \bibfield  {author} {\bibinfo {author} {\bibfnamefont {M.}~\bibnamefont
  {Tokuyama}}\ and\ \bibinfo {author} {\bibfnamefont {K.}~\bibnamefont
  {Kawasaki}},\ }\bibfield  {title} {\bibinfo {title} {Fractal dimensions for
  diffusion-limited aggregation},\ }\href
  {https://doi.org/https://doi.org/10.1016/0375-9601(84)91083-1} {\bibfield
  {journal} {\bibinfo  {journal} {Physics Letters A}\ }\textbf {\bibinfo
  {volume} {100}},\ \bibinfo {pages} {337} (\bibinfo {year}
  {1984})}\BibitemShut {NoStop}%
\bibitem [{\citenamefont {Matsushita}\ and\ \citenamefont
  {Fujikawa}(1990)}]{1990_matsushita_phyA}%
  \BibitemOpen
  \bibfield  {author} {\bibinfo {author} {\bibfnamefont {M.}~\bibnamefont
  {Matsushita}}\ and\ \bibinfo {author} {\bibfnamefont {H.}~\bibnamefont
  {Fujikawa}},\ }\bibfield  {title} {\bibinfo {title} {Diffusion-limited growth
  in bacterial colony formation},\ }\href@noop {} {\bibfield  {journal}
  {\bibinfo  {journal} {Physica A: Statistical Mechanics and its Applications}\
  }\textbf {\bibinfo {volume} {168}},\ \bibinfo {pages} {498} (\bibinfo {year}
  {1990})}\BibitemShut {NoStop}%
\bibitem [{\citenamefont {Tronnolone}\ \emph {et~al.}(2018)\citenamefont
  {Tronnolone}, \citenamefont {Tam}, \citenamefont {Szenczi}, \citenamefont
  {Green}, \citenamefont {Balasuriya}, \citenamefont {Tek}, \citenamefont
  {Gardner}, \citenamefont {Sundstrom}, \citenamefont {Jiranek}, \citenamefont
  {Oliver},\ and\ \citenamefont {Binder}}]{Tronnolone2018}%
  \BibitemOpen
  \bibfield  {author} {\bibinfo {author} {\bibfnamefont {H.}~\bibnamefont
  {Tronnolone}}, \bibinfo {author} {\bibfnamefont {A.}~\bibnamefont {Tam}},
  \bibinfo {author} {\bibfnamefont {Z.}~\bibnamefont {Szenczi}}, \bibinfo
  {author} {\bibfnamefont {J.~E.~F.}\ \bibnamefont {Green}}, \bibinfo {author}
  {\bibfnamefont {S.}~\bibnamefont {Balasuriya}}, \bibinfo {author}
  {\bibfnamefont {E.~L.}\ \bibnamefont {Tek}}, \bibinfo {author} {\bibfnamefont
  {J.~M.}\ \bibnamefont {Gardner}}, \bibinfo {author} {\bibfnamefont {J.~F.}\
  \bibnamefont {Sundstrom}}, \bibinfo {author} {\bibfnamefont {V.}~\bibnamefont
  {Jiranek}}, \bibinfo {author} {\bibfnamefont {S.~G.}\ \bibnamefont
  {Oliver}},\ and\ \bibinfo {author} {\bibfnamefont {B.~J.}\ \bibnamefont
  {Binder}},\ }\bibfield  {title} {\bibinfo {title} {Diffusion-limited growth
  of microbial colonies},\ }\href@noop {} {\bibfield  {journal} {\bibinfo
  {journal} {Scientific Reports}\ }\textbf {\bibinfo {volume} {8}},\ \bibinfo
  {pages} {5992} (\bibinfo {year} {2018})}\BibitemShut {NoStop}%
\bibitem [{\citenamefont {Brady}\ and\ \citenamefont {Ball}(1984)}]{Brady1984}%
  \BibitemOpen
  \bibfield  {author} {\bibinfo {author} {\bibfnamefont {R.~M.}\ \bibnamefont
  {Brady}}\ and\ \bibinfo {author} {\bibfnamefont {R.~C.}\ \bibnamefont
  {Ball}},\ }\bibfield  {title} {\bibinfo {title} {Fractal growth of copper
  electrodeposits},\ }\href@noop {} {\bibfield  {journal} {\bibinfo  {journal}
  {Nature}\ }\textbf {\bibinfo {volume} {309}},\ \bibinfo {pages} {225}
  (\bibinfo {year} {1984})}\BibitemShut {NoStop}%
\bibitem [{\citenamefont {Meakin}(1983)}]{Meakin1983}%
  \BibitemOpen
  \bibfield  {author} {\bibinfo {author} {\bibfnamefont {P.}~\bibnamefont
  {Meakin}},\ }\bibfield  {title} {\bibinfo {title} {Diffusion-controlled
  cluster formation in 2---6-dimensional space},\ }\href
  {https://doi.org/10.1103/PhysRevA.27.1495} {\bibfield  {journal} {\bibinfo
  {journal} {Phys. Rev. A}\ }\textbf {\bibinfo {volume} {27}},\ \bibinfo
  {pages} {1495} (\bibinfo {year} {1983})}\BibitemShut {NoStop}%
\bibitem [{\citenamefont {Sander}(2000)}]{Sander2000}%
  \BibitemOpen
  \bibfield  {author} {\bibinfo {author} {\bibfnamefont {L.~M.}\ \bibnamefont
  {Sander}},\ }\bibfield  {title} {\bibinfo {title} {Diffusion-limited
  aggregation: A kinetic critical phenomenon?},\ }\href
  {https://doi.org/10.1080/001075100409698} {\bibfield  {journal} {\bibinfo
  {journal} {Contemporary Physics}\ }\textbf {\bibinfo {volume} {41}},\
  \bibinfo {pages} {203} (\bibinfo {year} {2000})}\BibitemShut {NoStop}%
\bibitem [{\citenamefont {Pasinetti}\ \emph {et~al.}(2019)\citenamefont
  {Pasinetti}, \citenamefont {Ramirez}, \citenamefont {Centres}, \citenamefont
  {Ramirez-Pastor},\ and\ \citenamefont {Cwilich}}]{Pasinetti2019}%
  \BibitemOpen
  \bibfield  {author} {\bibinfo {author} {\bibfnamefont {P.~M.}\ \bibnamefont
  {Pasinetti}}, \bibinfo {author} {\bibfnamefont {L.~S.}\ \bibnamefont
  {Ramirez}}, \bibinfo {author} {\bibfnamefont {P.~M.}\ \bibnamefont
  {Centres}}, \bibinfo {author} {\bibfnamefont {A.~J.}\ \bibnamefont
  {Ramirez-Pastor}},\ and\ \bibinfo {author} {\bibfnamefont {G.~A.}\
  \bibnamefont {Cwilich}},\ }\bibfield  {title} {\bibinfo {title} {Random
  sequential adsorption on euclidean, fractal, and random lattices},\ }\href
  {https://doi.org/10.1103/PhysRevE.100.052114} {\bibfield  {journal} {\bibinfo
   {journal} {Phys. Rev. E}\ }\textbf {\bibinfo {volume} {100}},\ \bibinfo
  {pages} {052114} (\bibinfo {year} {2019})}\BibitemShut {NoStop}%
\bibitem [{\citenamefont {Ramirez}\ \emph {et~al.}(2019)\citenamefont
  {Ramirez}, \citenamefont {Centres},\ and\ \citenamefont
  {Ramirez-Pastor}}]{Ramirez_defect_2019}%
  \BibitemOpen
  \bibfield  {author} {\bibinfo {author} {\bibfnamefont {L.~S.}\ \bibnamefont
  {Ramirez}}, \bibinfo {author} {\bibfnamefont {P.~M.}\ \bibnamefont
  {Centres}},\ and\ \bibinfo {author} {\bibfnamefont {A.~J.}\ \bibnamefont
  {Ramirez-Pastor}},\ }\bibfield  {title} {\bibinfo {title} {Inverse
  percolation by removing straight rigid rods from square lattices in the
  presence of impurities},\ }\href {https://doi.org/10.1088/1742-5468/ab054d}
  {\bibfield  {journal} {\bibinfo  {journal} {Journal of Statistical Mechanics:
  Theory and Experiment}\ }\textbf {\bibinfo {volume} {2019}},\ \bibinfo
  {pages} {033207} (\bibinfo {year} {2019})}\BibitemShut {NoStop}%
\bibitem [{\citenamefont {Kundu}\ and\ \citenamefont
  {Mandal}(2021)}]{Kundu2021}%
  \BibitemOpen
  \bibfield  {author} {\bibinfo {author} {\bibfnamefont {S.}~\bibnamefont
  {Kundu}}\ and\ \bibinfo {author} {\bibfnamefont {D.}~\bibnamefont {Mandal}},\
  }\bibfield  {title} {\bibinfo {title} {Breaking universality in random
  sequential adsorption on a square lattice with long-range correlated
  defects},\ }\href {https://doi.org/10.1103/PhysRevE.103.042134} {\bibfield
  {journal} {\bibinfo  {journal} {Phys. Rev. E}\ }\textbf {\bibinfo {volume}
  {103}},\ \bibinfo {pages} {042134} (\bibinfo {year} {2021})}\BibitemShut
  {NoStop}%
\bibitem [{\citenamefont {Fusco}\ \emph {et~al.}(2021)\citenamefont {Fusco},
  \citenamefont {Tran-Phu}, \citenamefont {Cembran}, \citenamefont {Kiy},
  \citenamefont {Kluth}, \citenamefont {Nisbet},\ and\ \citenamefont
  {Tricoli}}]{Fusco2021}%
  \BibitemOpen
  \bibfield  {author} {\bibinfo {author} {\bibfnamefont {Z.}~\bibnamefont
  {Fusco}}, \bibinfo {author} {\bibfnamefont {T.}~\bibnamefont {Tran-Phu}},
  \bibinfo {author} {\bibfnamefont {A.}~\bibnamefont {Cembran}}, \bibinfo
  {author} {\bibfnamefont {A.}~\bibnamefont {Kiy}}, \bibinfo {author}
  {\bibfnamefont {P.}~\bibnamefont {Kluth}}, \bibinfo {author} {\bibfnamefont
  {D.}~\bibnamefont {Nisbet}},\ and\ \bibinfo {author} {\bibfnamefont
  {A.}~\bibnamefont {Tricoli}},\ }\bibfield  {title} {\bibinfo {title}
  {Engineering fractal photonic metamaterials by stochastic self-assembly of
  nanoparticles},\ }\href
  {https://doi.org/https://doi.org/10.1002/adpr.202100020} {\bibfield
  {journal} {\bibinfo  {journal} {Advanced Photonics Research}\ }\textbf
  {\bibinfo {volume} {2}},\ \bibinfo {pages} {2100020} (\bibinfo {year}
  {2021})}\BibitemShut {NoStop}%
\end{thebibliography}

\end{document}